\documentclass[default]{aastex701}

\usepackage{graphicx}

\newcommand\kms{km\,s$^{-1}$}
\shorttitle{}
\shortauthors{Nhung et al.}
\graphicspath{{./}{}}
\nolinenumbers
\begin{document}

\title{From the light curves of Long Period Variables to their evolution along the AGB}

\author[0000-0002-0311-0809]{Pham Tuyet Nhung}
\affiliation{Vietnam National Space Center (VNSC), Vietnam Academy of Science and Technology (VAST),\\
  18 Hoang Quoc Viet, Nghia Do, Ha Noi, Vietnam}
\email[show]{pttnhung@vnsc.org.vn}

\author{Mai Nhat Tan}
\affiliation{Vietnam National Space Center (VNSC), Vietnam Academy of Science and Technology (VAST),\\
18 Hoang Quoc Viet, Nghia Do, Ha Noi, Vietnam}
\email[show]{mainhattandhpt@gmail.com}

\author[0000-0002-3816-4735]{Do Thi Hoai}
\affiliation{Vietnam National Space Center (VNSC), Vietnam Academy of Science and Technology (VAST),\\
18 Hoang Quoc Viet, Nghia Do, Ha Noi, Vietnam}
\email[show]{dthoai@vnsc.org.vn}

\author[0000-0002-8979-6898]{Pierre Darriulat}
\affiliation{Vietnam National Space Center (VNSC), Vietnam Academy of Science and Technology (VAST),\\
18 Hoang Quoc Viet, Nghia Do, Ha Noi, Vietnam}
\email[show]{darriulat@vnsc.org.vn}


%% encouraged to surround collaboration identifiers with ()s. The af

\begin{abstract}

A sample of 116 Long Period Variables is studied with the aim to reveal relations between the properties of their light curves and of
their evolution along the Asymptotic Giant Branch (AGB). Each light curve is carefully scrutinized and its properties are summarized
in a set of five parameters. One of these, which measures the relative width of the light maxima and minima, is found to be
  particularly efficient at revealing significant correlations with the current state of the star on the AGB.
The picture that had been sketched in an earlier work is clarified and new features are revealed. In
particular, the evolution on the AGB of stars having not yet experienced a third dredge-up event is shown to go together with a closely
related path followed by the light curve in the parameter space. Moreover, evidence is found for the existence of two different types of
light curves for stars having experienced dredge-up events strong enough for their impact to be detectable.
For some, probably associated with higher initial mass stars, the light maximum becomes broader as the star evolves along the AGB,
  while for others it becomes narrower. Interpretations are
proposed, however often too speculative to be firmly ascertained: the results presented raise several unanswered questions and point to
the need for new observations and analyses, suggesting that light curves still carry more messages and information than has been
currently possible to decrypt.

\end{abstract}



\section{Introduction} \label{sec1}

In a recent article \citep{Hoai2025} (Paper I), we presented a detailed study of a sample of light curves of Mira variables selected for a
high density of observations, from which we inferred a number of relations between their properties and the evolution of the star along
the Asymptotic Giant Branch (AGB). In its wake, the present article clarifies the picture presented by Paper I and addresses questions
that were left unanswered by using a larger sample of curves and introducing new parameters.

We refer the reader to Paper I for an introduction to the physics of variable stars in general and of Mira variables in
particular. We simply recall that the basic mechanism causing stars to pulsate is the action of a valve allowing or preventing heat to
escape, depending on the state of the star. In the case of Cepheids, which were studied first, the valve is essentially the result of the
dependence of the opacity of an ionized helium layer on temperature and density (kappa mechanism). In the case of AGB stars
\citep{Herwig2005}, the energy transport is dominantly convective rather than radiative, which requires a dynamic theory of convection.
Recently, significant progress toward producing realistic three-dimensional models of convection has been achieved (\citealt{Addari2024} and references therein; \citealt{Freytag2023} and references therein)
and the role played by convective cells in the evolution of
AGB stars has become an important topic of study, both for observations and for modelling. However, as remarked by \citet{Buchler1993}
in the case of Cepheids, numerical simulations, the brute-force method that produces the most detailed and accurate description of
observations, fail to reveal simply the underlying mechanism at stake. This is still true today and the same can be said, even more
strongly, of Long Period Variables.

Fifty years ago, \citet{Schwarzschild1975} argued that convection was probably a major contributor to the mass loss rate of Red
Giants and was likely to materialize in the form of a very few convective cells, each covering typically a steradian. His argument was
based on a scaling of the Sun's supergranulation. Today's understanding of solar supergranulation has somewhat progressed \citep{Rincon2018},
but the argument still applies in broad terms and observations using VLTI and ALMA, together with VLBI
detections of masers, have confirmed the existence of such convective cells. In broad terms, it seems well established that the mass
loss of oxygen-rich stars has its source in radial movements associated with pulsations and the shock waves which they generate, both
at the global star and convective cell levels, whether at steradian scale with life times of years or at smaller scales with much shorter
life times \citep{Paladini2018, RosalesGuzman2024, Vlemmings2017, Vlemmings2019, Vlemmings2024, Darriulat2024}.

But there is more to it. The study of stellar pulsations offers a number of puzzling features that contribute to the frustration of
often failing to produce a simple description of what is observed. Much light was shed on these issues during the last quarter of the
past century. By then \citet{Lorenz1964} had studied the logistic map, a recurrence relation of the form $x_{n+1}=\lambda x_n(1-x_n)$ and understood how
simple non-linear dynamical equations can produce chaos; \citet{May1976} had realized the strong impact that such results can have on
many branches of physical sciences and several authors, such as Alexander Sharkovsky and Mitchell Feigenbaum had studied periodic
non-linear dynamical systems \citep{Feigenbaum1978}. In astrophysics, during the fourth quarter of the past century, such results were
applied to stellar pulsations, mostly Cepheids \citep{Buchler1993}. Starting with writing the complicated set of perturbative partial
differential equations describing the sources of non-linearity governing the physics of stellar pulsations, one can reduce it to a small
number $p$ (called number of principal modes) of ordinary differential equations. In the same way as mathematicians had shown how
simple mathematics can produce very complicated dynamics, astrophysicists have then shown how complex systems can be described
in simple terms. Examples are a non-resonant RR Lyrae model, quite accurately described by the system of two non-resonant
amplitude equations for the fundamental mode and the first overtone \citep{Buchler1987}; the 2:1 resonance between the
fundamental mode and the second overtone found to be responsible for the Hertzsprung Progression in bump Cepheids \citep{Hertzsprung1926, Kovacs1989};
or the 3:2 resonance found to produce alternating cycles and the 5:2 resonance being responsible for
period doubling. One may also mention two RV Tauri stars, with light curves displaying double-peaked oscillations and apparent
chaos: AC Her \citep{Kollath1998} and R Sct \citep{Buchler1995}.

Indeed, many peculiarities observed in the light curves of Long Period Variables are far from being their exclusive and
privileged property. Period-doubling is a common phenomenon in many branches of physics, humps and alternately deep and shallow
minima are often found in other types of stars. Examples are the above-mentioned RV Tauri stars, W Virginis stars \citep{Plachy2017},
BL Herculis stars \citep{Smolec2012, Soszynski2011}, RR Lyrae stars \citep{Smolec2015}, bump Cepheids \citep{Kovacs1989}, etc.
The dynamics of these stars are governed by different mechanisms, hydrogen or helium burning, radiative or
convective heat transfer, etc. and such peculiarities are simply general and well-known features of non-linear dynamical systems:
their interpretation in terms of specific underlying physical processes is not straightforward. When studying stellar pulsations, one must
keep in mind that these are complex dynamical systems, small changes in the parameters governing their dynamics may have major
impacts: the light curves are prone to experience strong and/or sudden changes in period, amplitude and regularity. In particular, when
seeking interpretations of the results of Fourier or wavelet analyses of light curves displaying complex behaviour, the remarkably
simple picture often offered by an analysis of the power spectrum should not be interpreted as evidence for pulsating stars being organ
pipes. This has been remarkably illustrated by \citep{Fokin1994}, who has produced a most insightful study of the pulsations of AC Her, an
RV Tauri star, showing how a sequence of complex events at stake inside the star could be simply described as a resonance between
fundamental and first overtone.

\section{The sample of light curves and their parameterization} \label{sec2}

In Paper I, we used two partly overlapping samples of curves, both selected from the list of Mira variables produced by \citet{MerchanBenitez2023}:
Sample A included 71 stars of spectral types C, S and M, the latter restricted to stars of which the spectrum had
been searched for technetium with a clear result, positive (referred to as Myes types) or negative (referred to as Mno types). Sample B
included 24 curves from sample A selected for a very high density of observations, allowing for a particularly reliable
parameterization; in addition, it included 8 curves of similar high quality, of stars of spectral type M of which the technetium content
is unknown.

The present sample includes 37 additional stars from the \citet{MerchanBenitez2023} selection, of which 7 chosen for
having a period in excess of 470 days, and two stars excluded from it as being classified as semi-regular, W Hya and RS Cyg, which
both display interesting features. W Hya was studied by \citet{Uttenthaler2011} as meander-type period-changing and was found to
display no technetium but some lithium in the spectrum, although unlikely to imply efficient Hot Bottom Burning (HBB) operation.
RS Cyg is a carbon-rich variable with a period of some 420 days and a low mass loss rate of 2$\times$10$^{-7}$ M$_\odot$yr$^{-1}$ \citep{Guandalini2006},
possibly presenting a double-peaking profile \citep{Cadmus2022, Cadmus2024}.

It is important to recall that, when evolving along the AGB, some stars pulsate, others don't. Among stars that pulsate, some
do it in a very regular way, with well-defined periods and oscillation amplitudes, others display instead significant irregularity, with
changing periods and amplitudes. The traditional separation between Mira variables and semi-regular variables is an attempt to
distinguish between them, however often failing to do so in a convincing way. In the present article, it is sufficient to be conscious that
during their AGB life, most stars spend some time displaying pulsations of sufficient regularity to allow for the definition of a mean
period and amplitude; we shall refer to these as being in the domain of stable pulsation; others do not, but there is continuity between
the two cases and several stars will be seen to be in a state of transition between the two domains.

We study light curves produced by the American Association of Variable Stars Observers (AAVSO), limited to the past
century, from January 1, 1925 to December 31, 2024, and to the visible band. This choice is dictated by the need to deal with curves
offering a large enough density of observations to allow for a reliable definition of their shapes. We inspected carefully and separately
each of the light curves associated with the sample of stars selected in the present study. Differences from cycle to cycle are often
stronger than between mean cycles of different stars, a well-known source of complexity that makes the study of such curves
particularly difficult. In particular, many different parameterizations may be devised in order to attempt as reliable and useful a
characterization of their shapes. In the present article, we use five such parameters, four of which were defined in Paper I: the mean
period $P$, the mean oscillation amplitude $A$, the mean asymmetry $\alpha$ between the ascending and descending branches and a parameter $\Delta$
measuring the irregularity of the curve. The fifth parameter, $W$, measures the width of the profile in each cycle and is defined below.
In addition to these five parameters chosen to characterize the light curve, we study their relation to two parameters that characterize
the evolution of the star along the AGB: a colour index C and the spectral type, M$_{i-j}$, S or C, both of which are listed in the \citet{MerchanBenitez2023}
list. We briefly comment on each of these below.

The period $P$ is measured in days. Small differences between the values listed by \citet{MerchanBenitez2023} and those
used in the present article or in Paper I are the result of their being evaluated on different intervals of time and are irrelevant to the
present study. Stronger period changes have been observed and studied in the published literature, \citet{Zijlstra2002a} being
first to make a distinction between different types of period changes, which has then been generally adopted in later studies \citep[][and references therein]{MerchanBenitez2023}.
R Cen and T UMi are commonly quoted as cases of extreme and sudden changes of the
periods and oscillation amplitudes of their light curves; we comment on these in Subsections 4.1 and 4.2, respectively. Other
examples of such changes, however less spectacular, are LX Cyg, BH Cru and RU Vul. LX Cyg and BH Cru have been shown by
\citet{Uttenthaler2011, Uttenthaler2016} to have recently experienced a third-dredge-up event and to have switched on this occasion from
spectral type S to C; they are part of the present sample. RU Vul is a semi-regular variable of spectral type M4, which has seen its
period decrease from $\sim$155 to $\sim$108 days in the past 55 years \citep{Templeton2008}.

Examples of major, but less sudden period changes are R Hya, R Aql and W Dra; they are part of the present sample. R Hya
saw its period decrease from $\sim$500 days to $\sim$360 days over 250 years. \citet{Wood1981} suggested that such decline was the result
of a recent thermal pulse, but \citet{Zijlstra2002b} raised questions in relation with its mass loss history and \citet{Uttenthaler2011}
considered the possibility of other scenarios. Recently, \citet{Joyce2024} and \citet{Fadeyev2024} have produced state-of-the-art models
of the pulsations; Fadeyev concludes that the best fit is obtained for an initial mass of 4.7 solar masses, and Joyce et al. obtain instead
an initial mass of 1.6 to 3.7 solar masses, illustrating the difficulty to produce reliable modelling.

In Paper I we introduced two parameters meant to measure the irregularity displayed by a light curve: $\Delta M_{\rm{max}}$ and $\Delta'M_{\rm{max}}$.
$\Delta M_{\rm{max}}$ is the rms deviation from its mean of the magnitude at maximum light, $\Delta'M_{\rm{max}}$ is the mean value of the difference of magnitude
between two successive light maxima. We studied the relation between these two parameters and gave evidence for different types of
irregularity. However, for the purpose of the arguments presented in the present study, it will be sufficient to use a single parameter,
the mean value $\Delta$ of $\Delta M_{\rm{max}}$ and $\Delta'M_{\rm{max}}$.

A particularly useful parameter largely used in Paper I is the asymmetry $\alpha=(T_{\rm{asc}}-T_{\rm{desc}})/(T_{\rm{asc}}+T_{\rm{desc}})$ between the durations of
the ascending and descending branches. In the case of sample B we preferred to use the equivalent parameter $\varphi_{\rm{min}}=T_{\rm{desc}}/P=\frac{1}{2}(1-\alpha)$.
Depending on the method used to measure such asymmetry, different systematic uncertainties are attached to $\varphi_{\rm{min}}$ and $\alpha$. In the present
article we systematically use the former. We recall that in Paper I (Sections 4.1 and 7.1), we noted important differences between the properties of the light
curves of spectral type Mno according to the value of $\varphi_{\rm{min}}$: we defined a first Mno family, Mno1, as having $\varphi_{\rm{min}}<0.57$. In contrast with
curves of the second family, Mno2, their light curves were found to display no hump on the ascending branch, to be more symmetric
with $\varphi_{\rm{min}}$ between 0.50 and 0.55 (compared with between 0.57 and 0.65 for Mno2), to have smaller periods, below 300 days, to be
more regular and to have smaller colour indices than Mno2. They sit on the left-hand side of the colour vs period diagrams first
produced by \citet{Uttenthaler2019}, while Mno2 occupies also the right-hand side. When observed in infrared, the oscillation
amplitude is less than 20\% of the value in the visible, while it exceeds 25\% for Mno2.

The mean oscillation amplitude displayed by the light curves, $A$, measured in units of magnitude, was evaluated for each light
curve of the Paper I sample, but had been made little use of. In the present study, we pay more attention to the information that it
carries.

\begin{figure*}
  \centering
  \includegraphics[width=5.5cm,trim={1cm 0 0cm 0},clip]{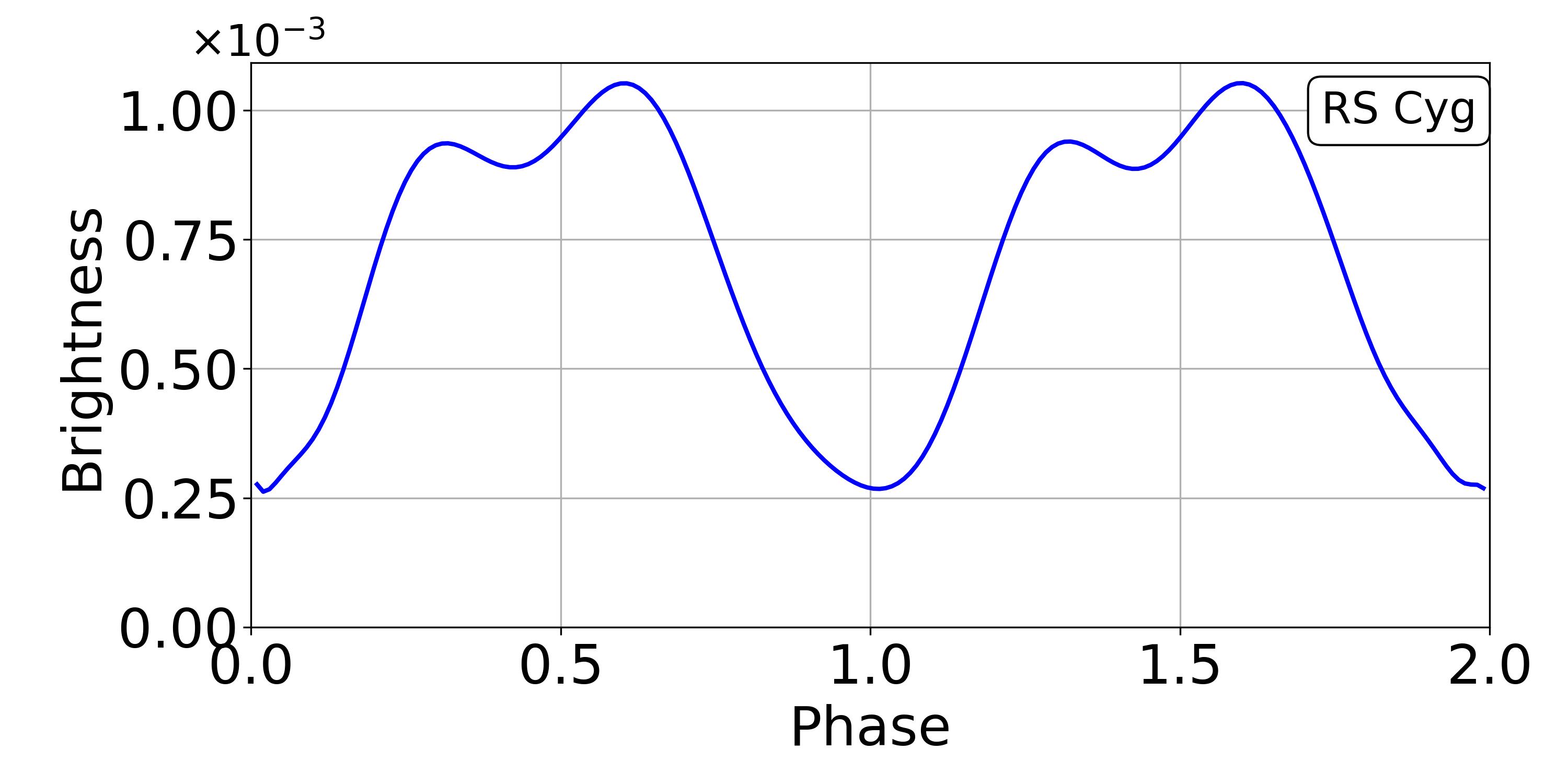}
  \includegraphics[width=5.5cm,trim={1cm 0 0cm 0},clip]{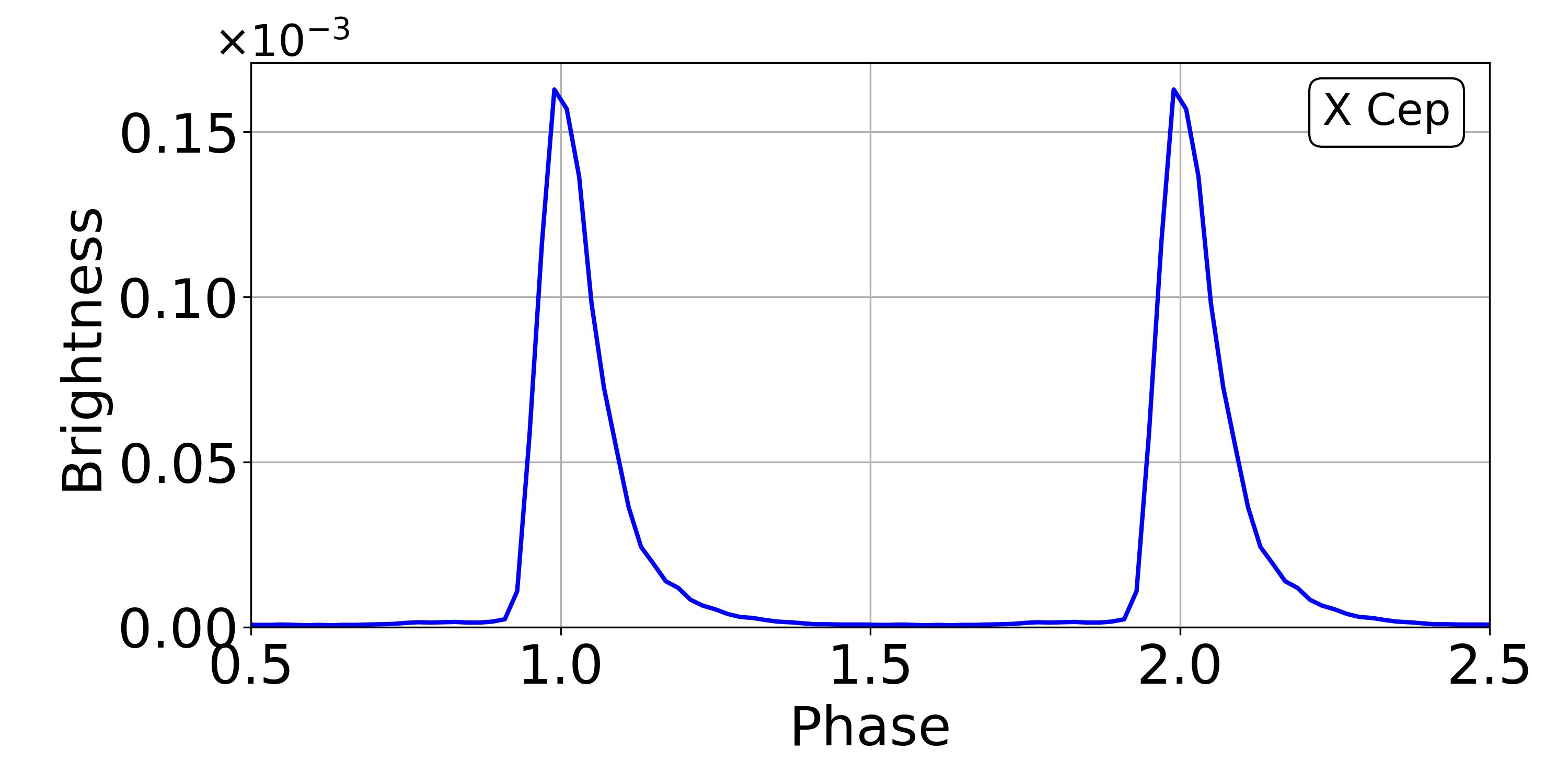}
  \includegraphics[width=6.7cm,trim={4cm 0 1cm 0},clip]{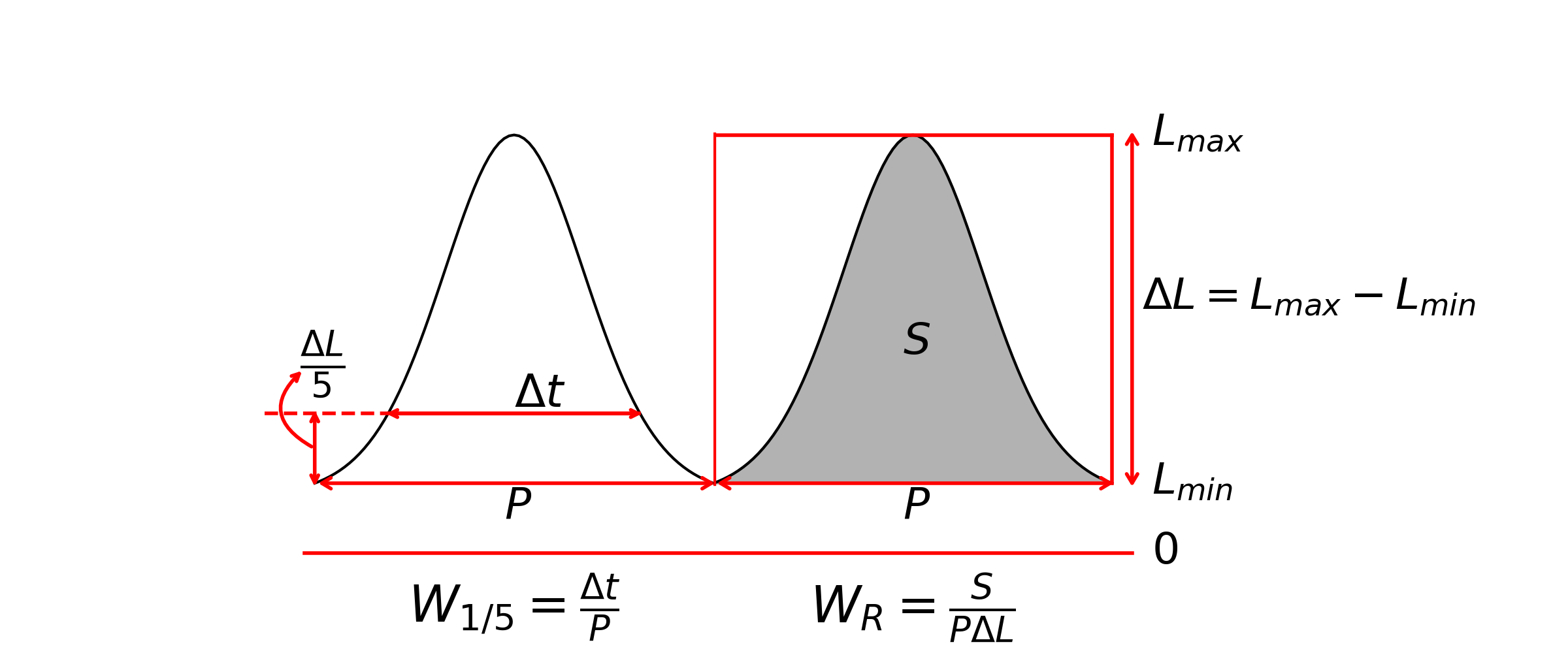}
  \caption{The left and central panels display extreme cases of broad (RS Cyg, $W$=0.66) and narrow (X Cep, $W$=0.16) profiles,
respectively. The right panel illustrates schematically the definition of parameters $W_{\rm{R}}$ and $W_{\rm{1/5}}$, of which $W$ is the geometric mean.}
 \label{fig1}
\end{figure*}

A well-known feature displayed by the light curves is a clear difference between profiles having a narrow maximum and a
broad minimum and profiles having a broad maximum and a narrow minimum. The left and central panels of Figure \ref{fig1} display two
extreme cases. Such distinction had already been made by \citet{Lebzelter2011}, however without producing very significant results. We
searched for parameters that quantify reliably this feature and found two good candidates, which we call $W_{\rm{1/5}}$ and $W_{\rm{R}}$ (Figure \ref{fig1}, right);
$W_{\rm{1/5}}$ is the width of the curve, relative to the period, at an apparent luminosity equal to 20\% of the luminosity amplitude of the
oscillation; $W_{\rm{R}}$ is the ratio between the luminosity measured from minimum light and integrated over a period and the product of its
value at maximum light by the period. Pulsations with relatively short values of the parameters mean a star that is mostly dim during
its cycle, pulsations with relatively large values mean a star that is mostly bright during its cycle. We find that, in general, the two
parameters are closely related, with, on average, $W_{\rm{1/5}}$$\sim$1.5 $W_{\rm{R}}$. Apart from a small difference between oxygen-rich
  and carbon-rich stars, we did not find significant cases of deviation between the information provided by the two parameters and considered
  it sufficient, for the purpose of the present study, to use a single width parameter $W$ that accounts for the sensitivity carried
    by each of $W_{\rm{1/5}}$ and $W_{\rm{R}}$. We define $W$ as the geometric mean of $W_{\rm{1/5}}$ and $W_{\rm{R}}$, $W=\sqrt{W_{\rm{1/5}}\times W_{\rm{R}}}$.
 The main difference between parameter $W$ and similar parameters commonly used in the published literature,
including Paper I, is its being evaluated from the luminosity rather than from the magnitude. This has a trivial but strong effect when
the oscillation amplitude is large; a good illustration is the case of $\chi$ Cyg which has a very low value of $W$, 0.18, while its profile,
studied in Paper I, is broad and displays a hump on the ascending branch.

The spectral type, M$_{i-j}$, S or C, is a reliable tracer of the evolution of the star along the AGB. The transition from M to S and from S to C, when it occurs, reveals unambiguously the last steps of such evolution before leaving the AGB. It is associated with a drastic change of the nature of the outer atmosphere, from oxygen rich to carbon-rich, resulting in a dramatic modification of its dust content. An essential milestone in the M spectral type is the presence or absence of technetium that reveals reliably the transition from the E-AGB to an advanced state of TP-AGB experiencing strong enough TDU events. Unfortunately, only few stars of M spectral type have been searched for technetium and the difficulty of the observation often prevents a clear answer to be obtained.

Finally, we use the colour indices K-[22] and [3.4]-[22] listed by \citet{MerchanBenitez2023} to reveal the emission of dust
in the circumstellar envelope; the K band is sensitive to water vapour emission \citep{Perrin2015} and the WISE [3.4] band may
contain strong absorption features of HCN and/or C$_2$H$_2$ molecules \citep{Aoki1999}. \citet{MerchanBenitez2023} remarked that the
advantage of the [3.4] band is that it was measured at the same time as the [22] band but its disadvantage is that bright Mira variables
saturate some detector pixels. We studied the relation obeyed by these parameters and concluded that, for the purpose of the present
study, it is sufficient to use a single parameter, $C$$=$$\frac{1}{2}$(K+[3.4])-[22].

There is a broad spectrum of density of observations among the curves studied in the present article. As a result, the
uncertainties attached to the curve parameters cover a broad range. We dealt carefully with each case separately, implying, in several
cases, a selection of cycles for which we trusted that a reliable evaluation could be obtained. In particular, several curves are the result
of observations made from a limited range of latitudes; when the star is low in the sky, this implies that it can be reliably observed
during only part of the year; when the pulsation period is close to a year, the result is that for a long sequence of years a same part of a
cycle is observed, producing so-called seasonal gaps: for example, the curve may cover the ascending branch for a decade and the
descending branch for the next. We have good control over such difficulties and trust that the uncertainties attached to the curve
parameters are small enough to have no significant impact on the arguments and conclusions presented in the article. However, we
refrained from quoting uncertainties on the values of each of the measured parameters.

\section{Distributions of the light curves in the parameter space}\label{sec3}
\subsection{The Mno spectral type}\label{sec3.1}

As was discussed in Paper I, the Mno spectral type is expected to include stars that have recently entered the TP-AGB or, possibly,
that are still in the E-AGB and will soon do so. The latter are still burning helium around the degenerate C/O core, while the former
are usually burning hydrogen around the depleted helium shell; however, when this has produced enough helium ashes in the inter-shell layer,
they experience a brief thermal pulse, which, when strong enough, causes a third dredge-up (TDU) event.
Low initial mass stars never experience such TDU event and spend their entire TP-AGB life as Mno types; only higher initial mass stars eventually
switch from Mno to Myes spectral type.

\begin{figure*}
  \centering
  \includegraphics[width=18cm,trim={0cm 0 0cm 0},clip]{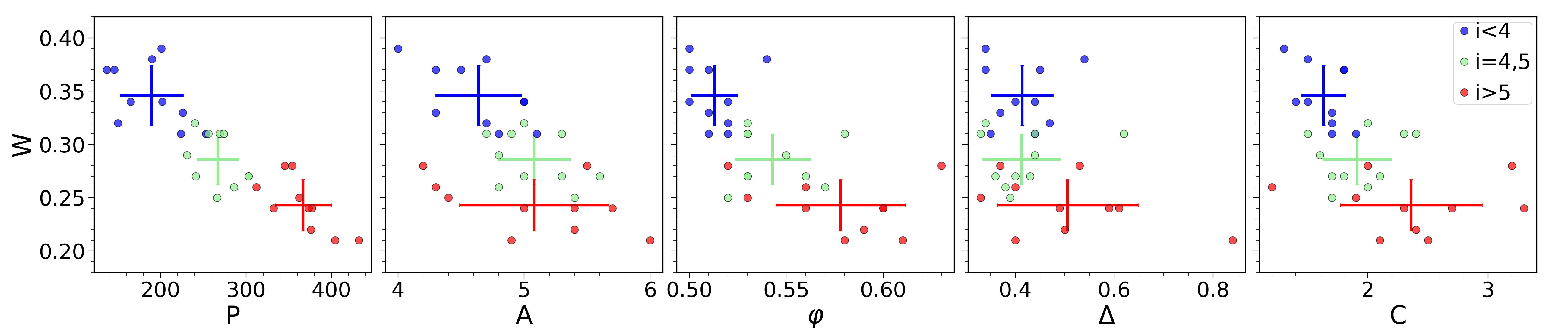}
  \caption{Distributions in parameter space of the light curves listed in Table \ref{tab1} (Mno). Projections on the $W$ vs $P$, $W$ vs $A$, $W$ vs $\varphi_{\rm{min}}$, $W$ vs $\Delta$
and $W$ vs $C$ planes are displayed from left to right. Different colours are used depending on the value of the index $i$ of the spectral type
M$i$-$j$ as shown in the inserts. The crosses show the mean values$\pm$rms deviations of the coordinates.}
 \label{fig2}
\end{figure*}

In the present section we study the light curves of stars having an Mno spectral type, as listed in Table \ref{tab1}. More precisely, we
exclude four such stars that display peculiarities that will be discussed in Subsection 4.2 (W Dra, W Hya, T Ari and U CMi), and we
include a star of spectral type M2, RT Cyg, for which technetium had been searched for and found unlikely to be present. Figure \ref{fig2}
illustrates the distributions of the curves in the parameter space by displaying projections on the $W$ vs $A$, $W$ vs $P$, $W$ vs $C$, $W$ vs $\varphi_{\rm{min}}$ and
$W$ vs $\Delta$ planes. It reveals a remarkable dependence on the temperature of the star, as measured by the value of index $i$ for an M$_{i-j}$
spectral type. As $i$ increases, namely as the star cools down, $W$ decreases and all other parameters increase. This evolution, for which
Table \ref{fig2} gives evidence, is particularly clear in the cases of $W$, $P$, $\varphi_{\rm{min}}$ and $C$, slightly less in the case of $A$ and unclear in the case of $\Delta$.

This result confirms and sheds light on the evolution from a first to a second family of curves that was suggested in Paper I.
However, it gives evidence for a continuous evolution rather than for a separation between two distinct families. We note that none of
the curves listed in Table \ref{tab1} displays a strong hump on the ascending branch. However, a few late types, such as R Cas, U Her, R LMi
and R Leo, do occasionally display a weak hump on the lower part of the ascending branch, in confirmation of the assignment of a \textbf{\textit{b}}
profile type made in Paper I.

This implies that as stars of the Mno spectral type evolve along the AGB, the width parameter of their light curves, $W$,
decreases. This raises two questions: where do they come from? And where will they go to?

As strongly suggested by the $W$ vs $P$ distribution shown in Figure \ref{fig2}, the progenitors of our Mno sample are variables of
smaller period, which are absent from the \citet{MerchanBenitez2023} selection because Mira variables are required by definition to
have periods in excess of 100 days. An interesting contribution to this issue is offered by a detailed study of the circumstellar envelope
of L$_2$ Pup by \citet{Kervella2014}, showing a strong similarity with R Vir, one of the early Mno stars in our sample, although their
spectral types are different: M3.5 for R Vir and M5 for L$_2$ Pup. Another very detailed study by \citet{Uttenthaler2024} has given evidence
for L$_2$ Pup to be Tc-poor and in an early stage of its evolution on the AGB, or even still on the RGB, approaching its tip. Numerous
observations at various wavelengths have been reported \citep[for a recent analysis see][and references therein]{Hoai2022a}, quoting a
low initial mass, 1 to 2 solar masses, and a very low mass loss rate, below $10^{-7}$ M$_\odot$yr$^{-1}$. L$_2$ Pup seems therefore to be a
good candidate for being in a state preceding that of the early Mno types in our sample, such as R Vir or R Vul. Its light curve
displays episodes during which the luminosity varies by less than one unit of magnitude, separated by episodes during which it
pulsates regularly ($\Delta$$\sim$0.2), with a mean oscillation amplitude of 1.4 units of magnitude, a stable period of 137 days and a width
parameter $W$$\sim$0.5. Its being in an unstable regime of pulsation at the limit of what was defined in the introduction as the domain of
stable pulsation, suggests that in their early life on the AGB, stars may see their light curves progressively evolve from irregular to
Mira-like once having reached the TP-AGB.

When the star has a large enough initial mass, it will at some point experience sufficiently strong TDU for its spectral type to
change to Myes and eventually S. To the extent that TDUs modify only progressively its outer atmosphere, we can expect the light curve of such a star to keep evolving in the direction of decreasing values of $W$. Yet, we see that while some curves of Myes and S type stars have indeed small enough $W$ values to be consistent with such a scenario, others have values of $W$ that are clearly too large, exceeding up to 0.4 in the case of R Cam and S UMa. This is surprising and requires an explanation. One may think of several reasons for the simple scenario to fail in such cases: the star may have left the Mno spectral type very early and its curve did not have time to evolve toward lower $W$ values but evolved instead toward larger $W$ values; or the curve of the star, when it had an Mno spectral type, was too irregular to be part of our sample; or it simply does not evolve as progressively as we assume... In any case, in order to clarify this important issue, it is convenient to discuss separately curves that may evolve toward smaller values of $W$ and curves that seem instead to evolve toward larger values of $W$. To this end, we define a value of $W$$=$0.3 to make the distinction between the two cases; this is of course an arbitrary choice, aimed only at easing the discussion, and should not be misinterpreted as a physically meaningful number.

\subsection{Oxygen-rich stars having possibly experienced strong TDU for which $W$$<$0.3}

\begin{figure*}
  \centering
  \includegraphics[width=18cm,trim={0cm 0 0cm 0},clip]{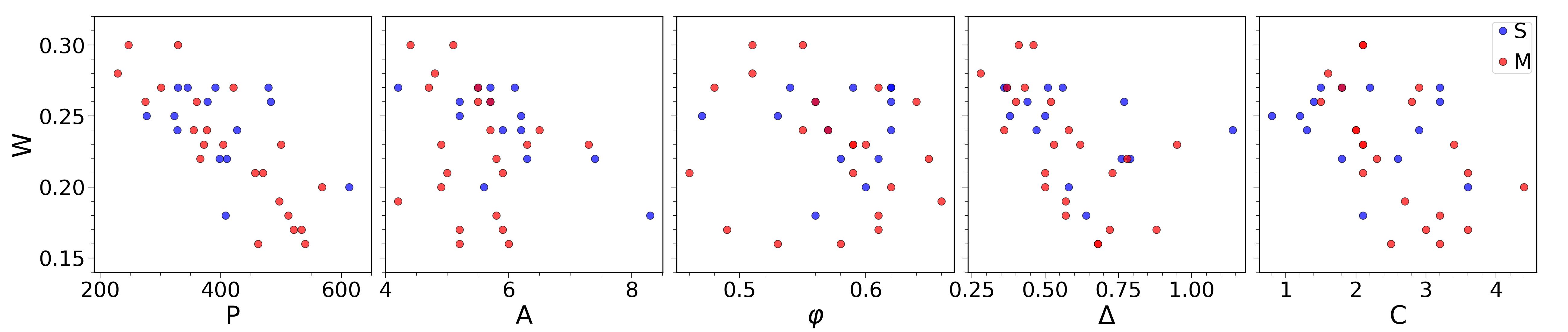}

  \caption{Distributions in parameter space of the light curves listed in Table \ref{tab3} ($W$$<$0.3). Projections on the $W$ vs $P$, $W$ vs $A$, $W$ vs $\varphi_{\rm{min}}$, $W$ vs $\Delta$
and $W$ vs $C$ planes are displayed from left to right. Different colours are used to distinguish between M (red) and S (blue) spectral types.}
 \label{fig3}
\end{figure*}

In line with the discussion presented in the preceding sub-section, we start with the study of the curves having $W$$<$0.3 of stars of M or S spectral type, excluding of course the Mno stars of Table \ref{tab1}. Their distribution in the parameter space is displayed in Figure \ref{fig3}. As can be seen in all panels of the figure, there is no clear distinction between the regions covered by S spectral types and by M spectral types. This is disappointing as it prevents using effectively the important milestone of the M to S transition as a tool to clarify the relation between the evolutions of the light curve in parameter space and of the star along the AGB. It suggests that such transition may occur at very different stages of the TP-AGB, depending on the star properties, mostly its initial mass.
Table \ref{tab3} lists all
curves of our sample, excluding the Mno curves listed in Table \ref{tab1}, that have an M or S spectral type and for which the light curve has a width
parameter $W$$<$0.3. Their distribution in the parameter space is displayed in Figure \ref{fig3}. As $W$ decreases, the period, the oscillation
amplitude and the colour index tend to increase. Remarkably, a dozen of curves display large irregularity parameters, well in excess of
0.6, in strong contrast with the Mno sample where a single curve, that of R Cas, has a large value of $\Delta$; it has also extreme values of $P$,
$A$ and $\varphi_{\rm{min}}$, suggesting that it is close to evolving to an Myes spectral type. The presence of a large group of irregular curves in the
present sample suggests that some are evolving toward a state of unstable pulsation and might be leaving the domain of stable
pulsation before becoming carbon-rich stars. Indeed, \citet{Schoier2001}, in a study of a large sample of bright carbon stars,
have shown that only a quarter of them pulsate as Mira variables.

The naive interpretation that the preceding subsection might have suggested, namely that stars would evolve across the Mno
sample, from $W$$\sim$0.4 down to $W$$\sim$0.2, before switching to the Myes, and later S, spectral types, is clearly flawed.
Thirteen curves, of which six of spectral type S, have $W$$>$0.25, larger than that of the $i$$>$5 sample of Figure \ref{fig2}.
This suggests therefore that when evolving
along the TP-AGB, stars will have experienced sufficient TDU at different stages of their evolution. This is indeed what should be
expected: stars of low initial masses spend their whole AGB life as Mno types and their curves may span down to $W$$\sim$0.2 as they
expand and become cooler. But stars with larger initial masses switch to the Myes spectral type earlier, before their curves reach low
values of $W$. Figure \ref{fig3} shows that S types appear already at $W$ values as large as 0.27 while Myes types reach down to $W$$=$0.16 and
large periods without switching to S types. In this context, we note that the seven curves that have been selected for having a large period, in excess of 470 days, are of the M spectral type without known Tc content and have indeed a low value of $W$, smaller than 0.22. The broad range of $W$ values covered by S type curves, namely by stars that are close to
becoming carbon stars, is challenging interpretation. As a consequence, the selection $W$$<$0.3, which we used to define the present
Table \ref{tab3} sample may be unjustified: some stars may leave the Mno spectral type when their curves have still a high value of $W$,
possibly as high as $\sim$0.4 in some cases. In order to progress in the understanding of this issue, it is therefore necessary to study the
sample of curves having $W$$>$0.3, which we do in the next subsection.

\begin{deluxetable*}{lcl ccc ccc c}
\tablenum{1}
\tablecaption{Parameters of the light curves of Mno spectral type studied in Subsection 3.1 \label{tab1}}
%\colnumbers
\tablewidth{0pt}
\tablehead{
\colhead{Name}&\colhead{$P$ (days)}&\colhead{Spctr}&\colhead{Tc}&\colhead{$^{12/13}$C}&\colhead{$C$ (mag)}&\colhead{$A$ (mag)}&\colhead{$W$}&\colhead{$\Delta$ (mag)}&\colhead{$\varphi_{\rm{min}}$}}
\startdata
RT Cyg	&190	&M2-8.8	&dbfl	&$-$	&1.5	&4.7	&0.38	&0.54	&0.54	\\
T Aqr	&202	&M2-5.5	&no	&$-$	&1.4	&5	&0.34	&0.4	&0.5	\\
X Peg	&201	&M2-5	&no	&$-$	&1.3	&4	&0.39	&0.34	&0.5	\\
T Her	&165	&M2.5-8	&no	&$-$	&1.5	&5	&0.34	&0.44	&0.52	\\
T Col	&226	&M3-6	&no	&$-$	&1.7	&4.3	&0.33	&0.37	&0.51	\\
T Eri	&253	&M3-5	&no	&$-$	&1.9	&4.8	&0.31	&0.44	&0.51	\\
RY Oph	&150	&M3-6	&no	&$-$	&1.7	&4.7	&0.32	&0.47	&0.52	\\
R Boo	&224	&M3-8	&no	&$-$	&1.7	&5.1	&0.31	&0.35	&0.52	\\
R Vul	&137	&M3-7	&no	&$-$	&1.8	&4.5	&0.37	&0.45	&0.51	\\
R Vir	&146	&M3.5-8.5	&no	&$-$	&1.8	&4.3	&0.37	&0.34	&0.5	\\
U Oct	&303	&M4-6	&no	&$-$	&1.8	&5.3	&0.27	&0.4	&0.53	\\
R Oph	&303	&M4-6	&no	&$-$	&2.1	&5.6	&0.27	&0.43	&0.56	\\
R Sgr	&269	&M4-6	&no	&$-$	&1.5	&4.9	&0.31	&0.44	&0.53	\\
RR UMa	&231	&M4	&no	&$-$	&1.6	&4.8	&0.29	&0.44	&0.55	\\
R Tri	&266	&M4-8	&no	&$-$	&1.7	&5.4	&0.25	&0.39	&0.52	\\
S Lac	&240	&M4-8.2	&no	&$-$	&2	&5	&0.32	&0.34	&0.53	\\
T UMa	&256	&M4-7	&no	&$-$	&2.4	&5.3	&0.31	&0.62	&0.58	\\
R Aql	&274	&M5-9	&no	&8	&2.3	&4.7	&0.31	&0.33	&0.53	\\
X CrB	&241	&M5-7	&no	&$-$	&1.7	&5	&0.27	&0.36	&0.53	\\
R Del	&286	&M5-6	&no	&$-$	&2	&4.8	&0.26	&0.38	&0.57	\\
R Cnc	&362	&M6-9	&no	&17	&1.9	&4.4	&0.25	&0.33	&0.53	\\
RU Hya	&332	&M6-8.8	&no	&$-$	&3.3	&5.4	&0.24	&0.59	&0.6	\\
R Peg	&377	&M6-9	&no	&10	&2.3	&5	&0.24	&0.49	&0.56	\\
W Peg	&345	&M6-8	&no 	&17	&2	&4.2	&0.28	&0.37	&0.52	\\
RS Vir	&354	&M6-8	&no	&$-$	&3.2	&5.5	&0.28	&0.53	&0.63	\\
R Cas	&432	&M6-10	&no	&12	&2.1	&6	&0.21	&0.84	&0.61	\\
U Her	&404	&M6.5-9.5	&no	&19	&2.5	&4.9	&0.21	&0.4	&0.58	\\
R LMi	&373	&M6.5-9.0	&no	&12	&2.7	&5.7	&0.24	&0.61	&0.6	\\
R Leo	&312	&M7-9	&no	&10	&1.2	&4.3	&0.26	&0.4	&0.56	\\
W Eri	&376	&M7-9	&no	&$-$	&2.4	&5.4	&0.22	&0.5	&0.59	\\
\enddata
\end{deluxetable*}
%%%

\begin{deluxetable*}{lccc}%{rrrrrrrrrrr}%
\tablenum{2}
\tablecaption{Mean values $\pm$ rms deviations from the mean of the parameters of the curves of Mno spectral type listed in Table \ref{tab1}. Three
different groups are considered depending on the index $i$ of spectral type M$_{i-j}$. \label{tab2}}
%\tablewidth{0pt}
\tablehead{\colhead{} &\colhead{$i$$<$4} &\colhead{$i$$=$4 or 5} &\colhead{$i$$>$5} }
\startdata
$<$$W$$>$$\pm$rms($W$)	& 0.35$\pm$0.03	& 0.29$\pm$0.02	& 0.24$\pm$0.03	\\
$<$$P$$>$$\pm$rms($P$) [days]	& 189$\pm$37	& 267$\pm$24	& 367$\pm$33	\\
$<$$A$$>$$\pm$rms($A$) [mag]	& 4.6$\pm$0.3	& 5.1$\pm$0.3	& 5.1$\pm$0.6	\\
$<$$\Delta$$>$$\pm$rms($\Delta$) [mag]	& 0.41$\pm$0.06	& 0.41$\pm$0.08	& 0.51$\pm$0.14	\\
$<$$\varphi_{\rm{min}}$$>$$\pm$rms($\varphi_{\rm{min}}$)	& 0.51$\pm$0.01	& 0.54$\pm$0.02	& 0.58$\pm$0.03	\\
$<$$C$$>$$\pm$rms($C$) [mag]	& 1.6$\pm$0.2	& 1.9$\pm$0.3	& 2.4$\pm$0.6
\enddata
\end{deluxetable*}

\subsection{Oxygen-rich stars having possibly experienced strong TDU for which $W$$>$0.3}

We now turn to the study of the curves having $W$$>$0.3 of stars of M or S spectral type, excluding of course the Mno stars of Table \ref{tab1}.
They are listed in Table \ref{tab4} and Figure \ref{fig4} displays their distribution in the parameter
space. At first glance, Figure \ref{fig4} seems to suggest that the $W$$>$0.3 curves are successors of the Mno$_{i-j}$ curves having $i$$<$4, while the
$W$$<$0.3 curves are successors of the Mno$_{i-j}$ curves having $i$$>$3; moreover, the location of the $W$$>$0.3 curves between the Mno and
carbon-rich samples may suggest that the stars having $W$$>$0.3 curves will evolve to the carbon-rich variables of the present sample. Yet, several distributions, in particular in the $W$ vs $\Delta$ and $W$ vs $C$ planes, show continuity between the Table \ref{tab3} ($W$$<$0.3) and Table \ref{tab4} ($W$$>$0.3) samples, raising questions on the opportunity to make a distinction between them. This is further illustrated in Section 5, where both samples are considered jointly.
A close inspection of the properties of the curves listed in Table \ref{tab4} is therefore necessary to understand whether such crude impressions
are or not justified.

The twelve curves having $W$$<$0.4 include the seven stars of spectral type M and five of the eight stars of spectral type S which
the sample contains and can be interpreted as successors of the Mno$_{i-j}$ curves having $i$$<$4: they have $<$$W$$>$=0.34 compared with 0.35,
$<$$P$$>$=334 compared with 189, $<$$A$$>$=4.7 compared with 4.6, $<$$\Delta$$>$=0.32 compared with 0.41, $<$$\varphi_{\rm{min}}$$>$=0.51 compared with
0.51 and $<$$C$$>$=1.4 compared with 1.6. Namely, while the period has much increased, and $\Delta$ a little decreased, the other parameters remained
nearly unchanged on average. This is compatible with a future evolution toward the sample of carbon-rich stars as long as it would
make room for a significant decrease of the oscillation amplitude.

This leaves three additional curves having $W$$>$0.4, all of stars of spectral type S: R Cam, T Cam and S UMa. They are part of
Sample B of Paper I. Such a small sample does not allow for drawing reliable conclusions. Yet, as can be seen on Figure \ref{fig4}, while the
periods are compatible with an evolution toward the sample of carbon-rich stars, the values of $C$ and $\Delta$ are surprisingly small and those
of $A$ surprisingly large.

\begin{deluxetable*}{lcl ccc ccc c}%{rrrrrrrrrrr}%
\tablenum{3}
\tablecaption{List of the light curves of M and S spectral types having $W$$<$0.3. The curves listed in Table \ref{tab1} are excluded. \label{tab3}}
%\tablewidth{0pt}
\tablehead{
  \colhead{Name}&\colhead{$P$ (days)}&\colhead{Spctr}&\colhead{Tc}&\colhead{$^{12/13}$C}&\colhead{$C$ (mag)}&\colhead{$A$ (mag)}&\colhead{$W$}&\colhead{$\Delta$ (mag)}&\colhead{$\varphi_{\rm{min}}$}}
\startdata
R And	&410	&S3.5-8.8	&yes	&40	&2.6	&7.4	&0.22	&0.76	&0.61	\\
W And	&398	&S6-9	&yes	&$-$	&1.8	&6.3	&0.22	&0.79	&0.58	\\
S Cas	&613	&S3-5	&$-$	&32	&3.6	&5.6	&0.2	&0.58	&0.6	\\
$\chi$ Cyg	&408	&S6-9	&yes	&36	&2.1	&8.3	&0.18	&0.64	&0.56	\\
R Cyg	&427	&S2.5-6	&yes	&29	&2.9	&6.2	&0.24	&1.14	&0.62	\\
U Cas	&277	&S3-8	&$-$	&$-$	&1.2	&6.2	&0.25	&0.38	&0.53	\\
W Aql	&483	&S3-6	&yes	&26	&3.2	&5.2	&0.26	&0.77	&0.62	\\
WY Cas	&479	&S6.5	&yes	&$-$	&3.2	&6.1	&0.27	&0.56	&0.62	\\
X And	&345	&S2-5	&$-$	&$-$	&2.2	&5.5	&0.27	&0.51	&0.62	\\
SZ Cep	&329	&S3.5-4	&$-$	&$-$	&1.8	&5.7	&0.27	&0.36	&0.59	\\
R Lyn	&378	&S2.5-6	&yes	&$-$	&1.4	&5.7	&0.26	&0.44	&0.56	\\
RR And	&328	&S6.5	&$-$	&$-$	&1.3	&5.9	&0.24	&0.47	&0.57	\\
S Cyg	&323	&S2.5	&$-$	&$-$	&0.8	&5.2	&0.25	&0.5	&0.47	\\
T Sgr	&391	&S4.5-5.5	&yes	&$-$	&1.5	&4.2	&0.27	&0.37	&0.54	\\
R Aur	&457	&M6.5-9.5	&yes	&33	&2.1	&5.9	&0.21	&0.5	&0.46	\\
RU Her	&486	&M6-9	&yes	&25	&2.3	&5.9	&0.18	&0.74	&0.54	\\
R Ser	&355	&M5-9	&yes	&14	&2	&6.5	&0.24	&0.58	&0.57	\\
U Ari	&372	&M4-7.5	&yes	&$-$	&2.1	&6.3	&0.23	&0.62	&0.59	\\
V CMi	&366	&M4-10	&yes	&$-$	&2.3	&5.8	&0.22	&0.78	&0.65	\\
R Hor	&404	&M5-8	&yes	&$-$	&2.1	&7.3	&0.23	&0.53	&0.59	\\
S Vir	&377	&M6-9.5	&yes	&$-$	&2	&5.7	&0.24	&0.36	&0.55	\\
V Gem	&275	&M4-8	&yes	&$-$	&1.5	&5.7	&0.26	&0.4	&0.56	\\
S Ori	&421	&M6.5-9.5	&yes	&45	&1.8	&4.7	&0.27	&0.37	&0.48	\\
V Cam	&521	&M7	&$-$	&$-$	&3.6	&5.9	&0.17	&0.72	&0.61	\\
X Cep	&540	&M4-7	&$-$	&$-$	&3.2	&6	&0.16	&0.68	&0.58	\\
UX Cyg	&568	&M4-6.5	&$-$	&$-$	&4.4	&4.9	&0.2	&0.5	&0.62	\\
V Del	&534	&M4-6	&$-$	&$-$	&3	&5.2	&0.17	&0.88	&0.49	\\
Z Pup	&512	&M4-9	&dbfl	&$-$	&3.2	&5.8	&0.18	&0.57	&0.61	\\
Z Cas	&497	&M7	&$-$	&$-$	&2.7	&4.2	&0.19	&0.57	&0.66	\\
RU Aur	&470	&M7-9	&$-$	&$-$	&3.6	&5	&0.21	&0.73	&0.59	\\
S CrB	&360	&M6-8	&poss	&$-$	&2.8	&5.5	&0.26	&0.52	&0.64	\\
R UMa	&301	&M3-9	&$-$	&$-$	&2.9	&5.5	&0.27	&0.43	&0.61	\\
CH Pup	&500	&M7	&$-$	&$-$	&3.4	&4.9	&0.23	&0.95	&0.6	\\
R CVn	&329	&M6-9	&poss	&$-$	&2.1	&4.4	&0.3	&0.41	&0.51	\\
V Cas	&229	&M5-8.5	&$-$	&$-$	&1.6	&4.8	&0.28	&0.28	&0.51	\\
R Dra	&247	&M5-9	&dbfl	&$-$	&2.1	&5.1	&0.3	&0.46	&0.55	\\
R Tel	&462	&M5-7	&$-$	&$-$	&2.5	&5.2	&0.16	&0.68	&0.53	\\
\enddata
\end{deluxetable*}

The trend toward low colour indices is already apparent in the Table \ref{tab4} sample of the twelve curves having $W$$<$0.4. Indeed, out
of the 548 curves selected by \citet{MerchanBenitez2023}, 40 have [3.4]-[22]$<$1.25, of which 19 are of S spectral type, 18 of M
spectral type and 3 of C spectral type. Of these, 10 are part of the Table \ref{tab4} sample and only 5, all of S spectral type with $W$ between
0.24 and 0.27, are part of the much larger Table \ref{tab3} sample. This is evidence for a group of stars of very low colour indices, mostly of S,
but also Myes spectral types, having curves of large $W$ parameter, to own a specific identity that must be associated with well-defined
physical properties. A likely interpretation is offered by \citet{Hashimoto1998} in the case of R Hya, a star of the Table \ref{tab4} sample having an M6-
9 spectral type, a colour index $C$$=$0.9 and a broad curve profile, $W$$=$0.36. The claim is that mass-loss has recently nearly stopped,
causing the circumstellar envelope to expand away from the star and become too cold for the silicate dust grains to significantly emit
in the infrared. Recent observations and analyses \citep{Decin2008, Baudry2023, Joyce2024, ZhaoGeisler2012}
have confirmed the validity of this interpretation and suggested that the cause of the absence of significant mass-loss is the likely
occurrence of a thermal pulse some 200 years ago. 

More generally, as remarked by \citet{Marigo2020}, low mass loss rates, and therefore low colour indices, are expected
when the C/O ratio is close to unity. In such a case, the amount of atmospheric oxygen when C/O$<$1 (S type) or of atmospheric carbon
when C/O$>$1 (C type) is too low to allow for significant formation of dust on which radiation pressure could produce the mass-losing
wind.

In summary, an interpretation of the Table \ref{tab4} curves as associated with large initial mass stars near the end of the TP-AGB,
having light curves evolving from early Mno type ($i$$<$4) to the carbon-rich Table \ref{tab5} sample is plausible but far from being firmly
established and raises several questions that deserve careful consideration. The strong evidence for a trend toward low colour indices when the star evolves on the AGB from M to S spectral types, is particularly puzzling in this respect: it goes against an evolution toward the curves of C spectral type, which have instead larger colour indices. The argument proposed by \citet{Marigo2020}, concerning the expected scarcity of dust when C/O approaches unity, is very relevant but raises questions concerning its detailed application, not only for the curves of Table \ref{tab4} but also for those of Table \ref{tab3}.

\begin{deluxetable*}{lcl ccc ccc c}%{rrrrrrrrrrr}%
\tablenum{4}
\tablecaption{List of curves of M and S spectral types having $W$$>$0.3. The curves listed in Table \ref{tab1} are excluded.\label{tab4}}
%\tablewidth{0pt}
\tablehead{
  \colhead{Name}&\colhead{$P$ (days)}&\colhead{Spctr}&\colhead{Tc}&\colhead{$^{12/13}$C}&\colhead{$C$ (mag)}&\colhead{$A$ (mag)}&\colhead{$W$}&\colhead{$\Delta$ (mag)}&\colhead{$\varphi_{\rm{min}}$}}
\startdata
R Cam	&270	&S2-8	&$-$	&$-$	&0.4	&4.5	&0.41	&0.28	&0.52	\\
T Cam	&376	&S4-8.5	&yes	&31	&0.7	&5.4	&0.4	&0.18	&0.49	\\
S UMa	&226	&S0-5	&yes	&$-$	&1.2	&3.7	&0.46	&0.09	&0.5	\\
V Cnc	&272	&S0-7	&yes	&$-$	&1.4	&5	&0.33	&0.3	&0.55	\\
FF Cyg	&328	&S6	&$-$	&$-$	&1	&4.5	&0.34	&0.22	&0.5	\\
Z Del	&305	&S5.2-7	&yes	&$-$	&1.4	&5.5	&0.3	&0.45	&0.51	\\
R Gem	&361	&S2-8	&yes	&22	&2	&6.2	&0.31	&0.52	&0.64	\\
T Gem	&287	&S1.5-9	&yes	&$-$	&0.5	&5.4	&0.38	&0.21	&0.5	\\
R Hya	&386	&M6-9	&yes	&26	&0.9	&3.7	&0.36	&0.38	&0.49	\\
T Cep	&338	&M5-9	&yes	&33	&1.2	&4.1	&0.33	&0.29	&0.47	\\
S Her	&307	&M4.8-7.5	&yes	&$-$	&1.1	&5.4	&0.31	&0.33	&0.49	\\
T Cas	&445	&M6-9	&prob	&33	&1.9	&4.3	&0.39	&0.27	&0.43	\\
U UMi	&324	&M6-8	&$-$	&$-$	&1.6	&3.3	&0.37	&0.24	&0.5	\\
S CMi	&333	&M6-8	&yes	&18	&1.9	&4.9	&0.32	&0.33	&0.49	\\
Z Peg	&326	&M6-8.5	&yes	&$-$	&1.7	&4.7	&0.3	&0.33	&0.5	\\
\enddata
\end{deluxetable*}

\subsection{Carbon-rich stars}

\begin{figure*}
  \centering
  \includegraphics[width=5cm,trim={0cm 0 1.5cm 0},clip]{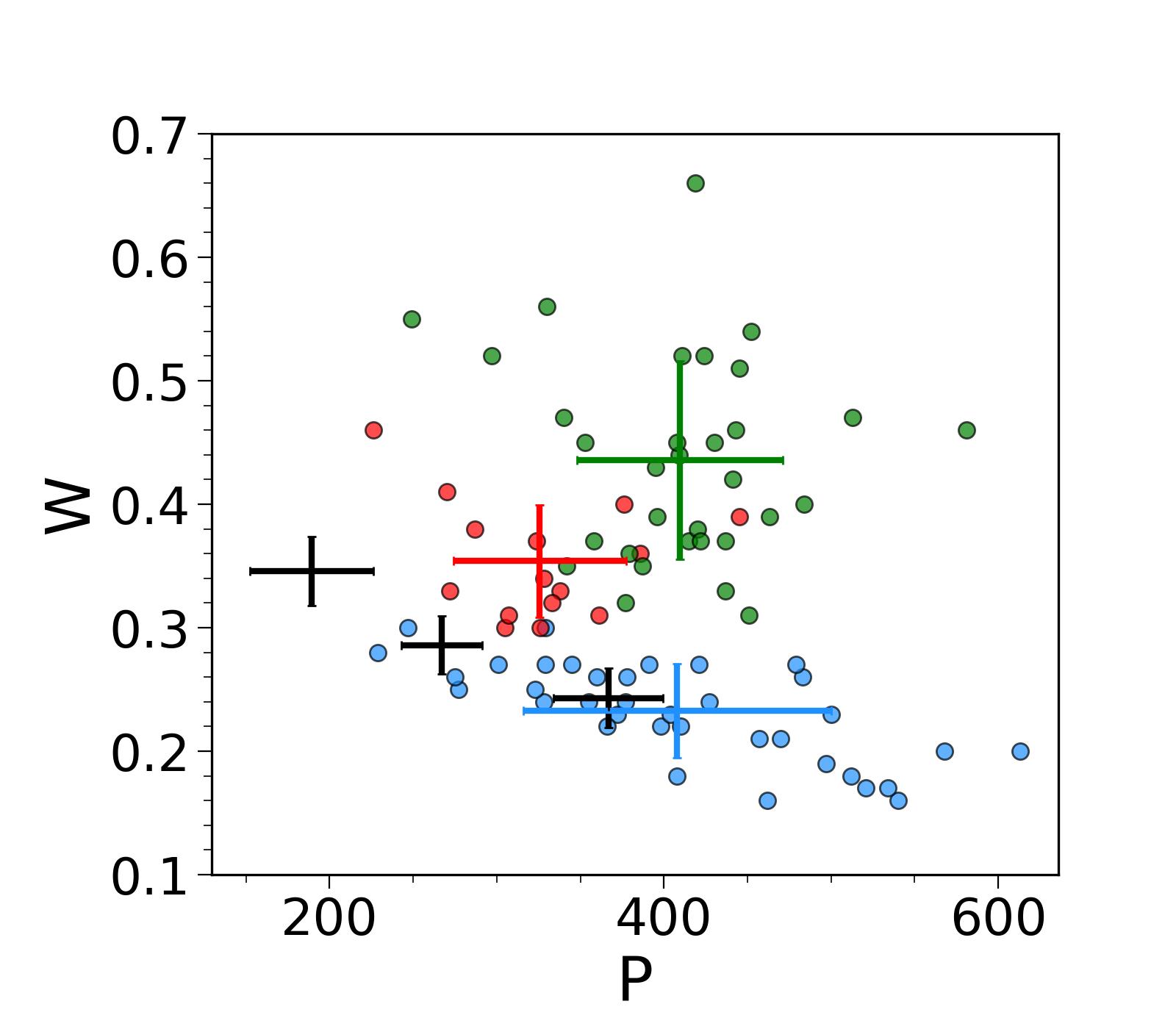}
\includegraphics[width=5cm,trim={0cm 0 1.5cm 0},clip]{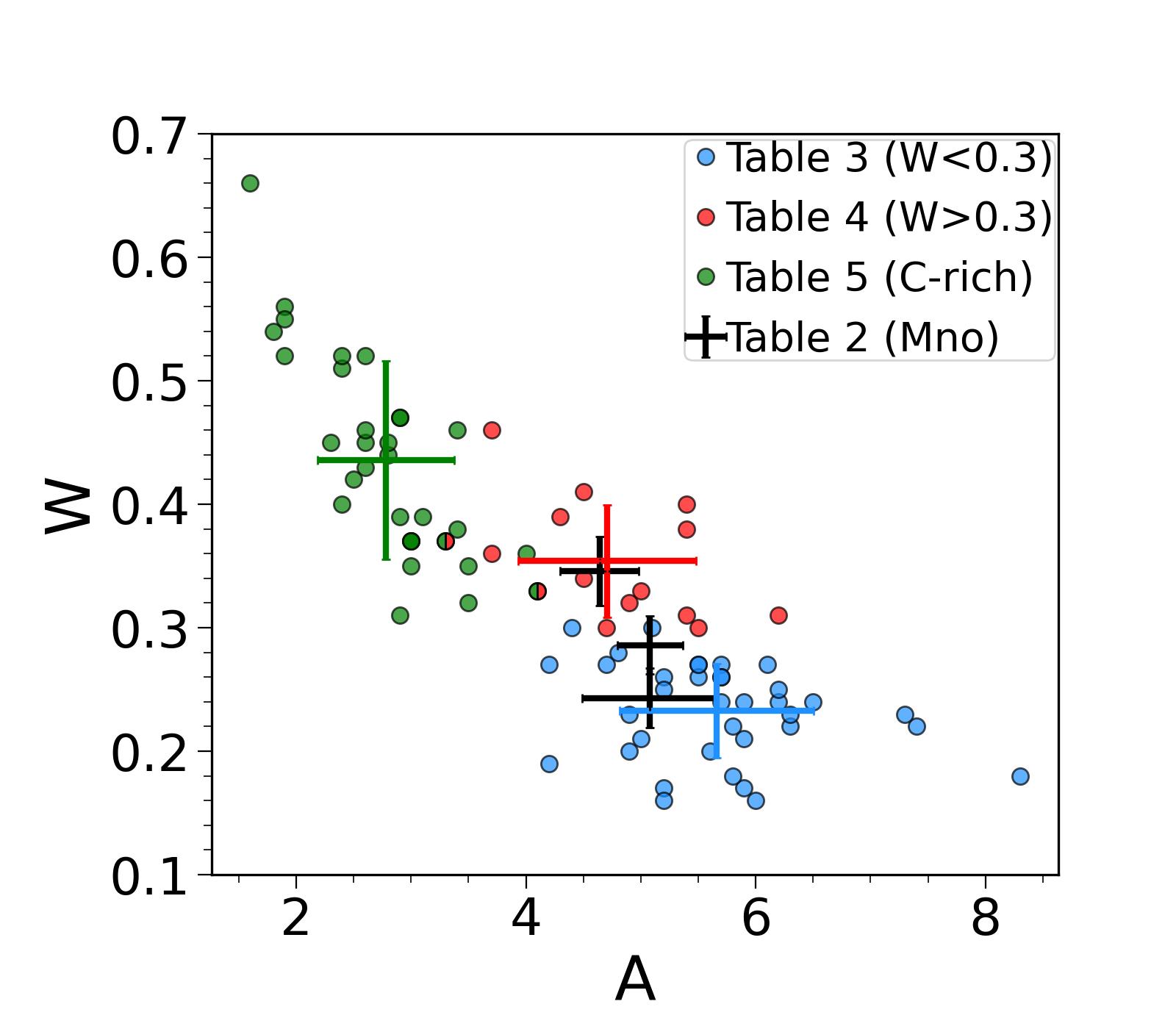}\\
\includegraphics[width=5cm,trim={0cm 0 1.5cm 0},clip]{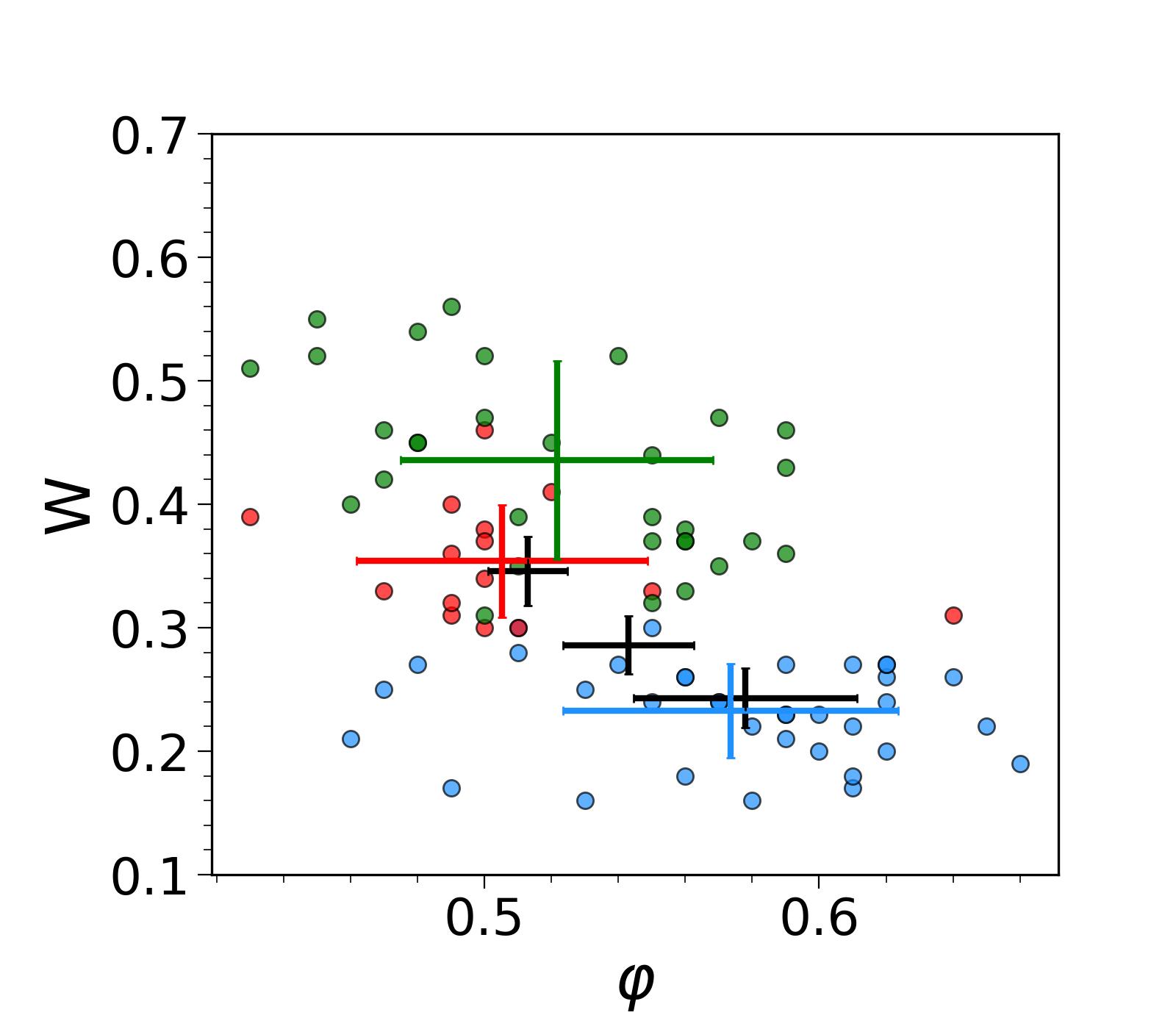}
\includegraphics[width=5cm,trim={0cm 0 1.5cm 0},clip]{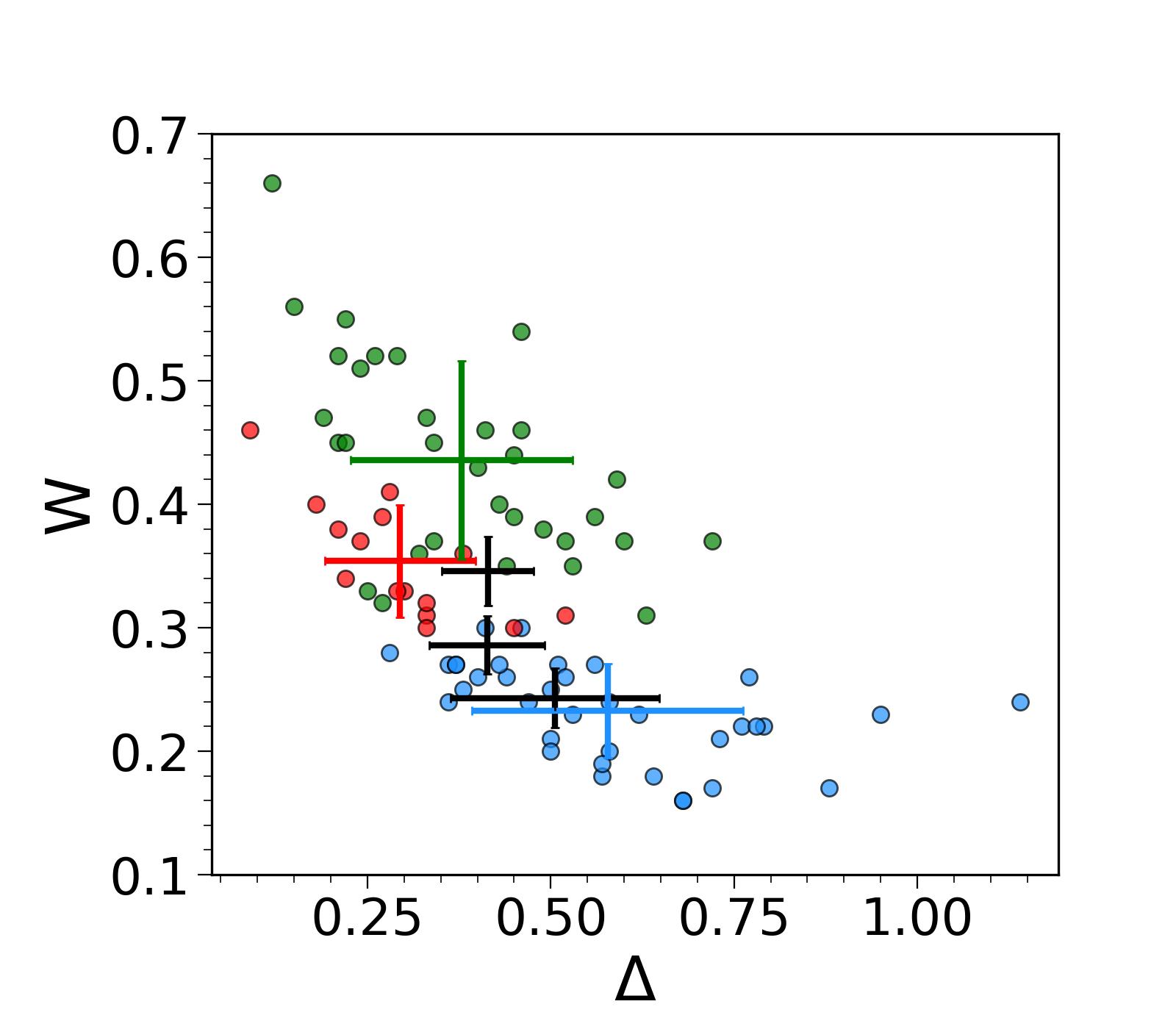}
\includegraphics[width=5cm,trim={0cm 0 1.5cm 0},clip]{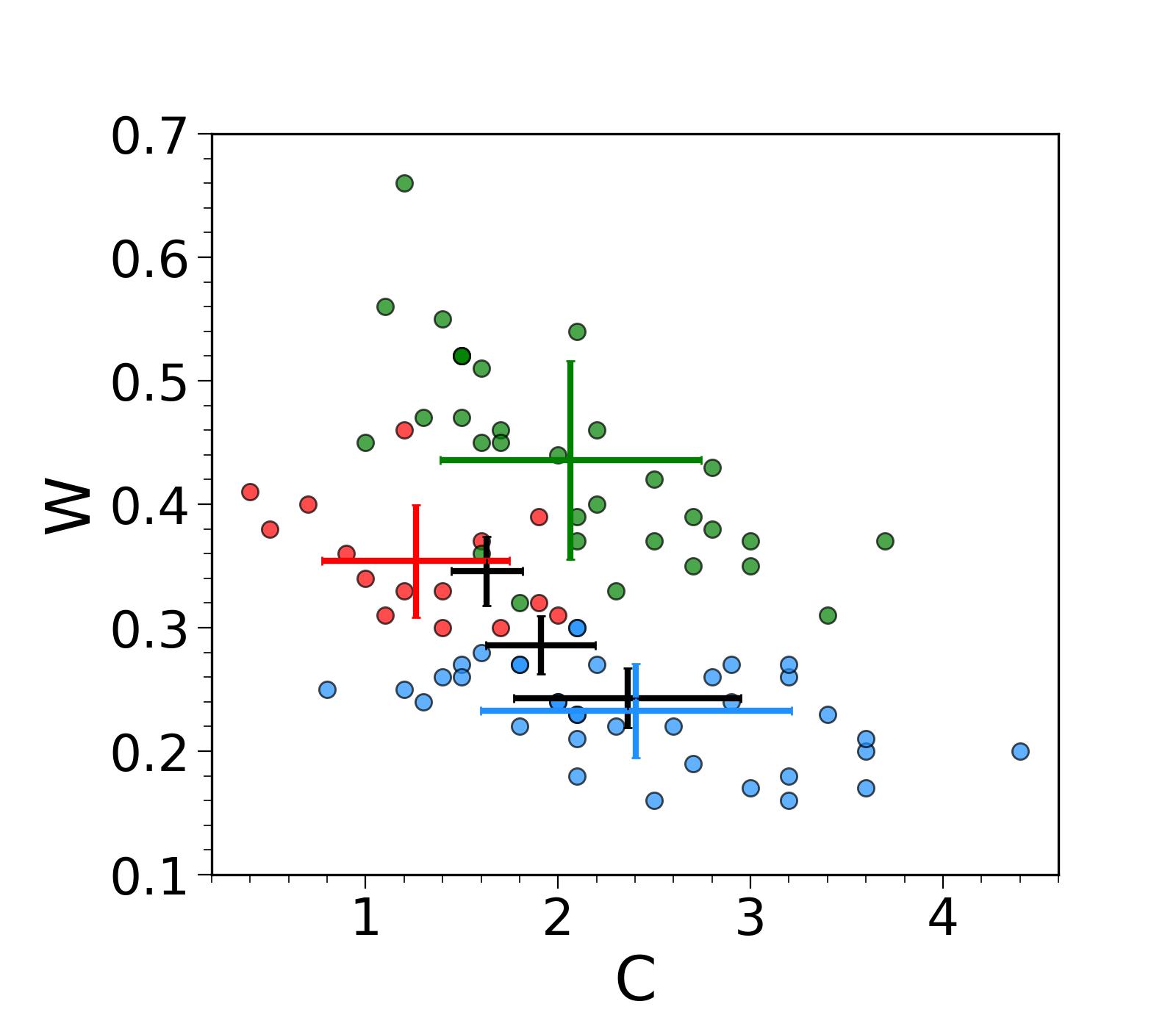}

  \caption{Distributions in parameter space of the light curves listed in Tables \ref{tab1} (Mno), \ref{tab3} ($W$$<$0.3), \ref{tab4} ($W$$>$0.3) and \ref{tab5} (C-rich). Black crosses show the light curves of
Table \ref{tab1} as averaged in Table \ref{tab2}. Coloured crosses show the mean$\pm$rms deviation of the coordinates for the samples of Tables \ref{tab3} to \ref{tab5}.
Projections on the $W$ vs $P$, $W$ vs $A$, $W$ vs $\varphi_{\rm{min}}$, $W$ vs $\Delta$ and $W$ vs $C$ planes are displayed from left to right. Different colours are used to
distinguish between the different samples as shown in the insert.}
 \label{fig4}
\end{figure*}

The light curves of the carbon-rich stars of the present sample are listed in Table \ref{tab5} and their distribution in the parameter space is
illustrated in Figure \ref{fig4}. The width parameter $W$ exceeds 0.3. A clear correlation is seen between $W$ and $A$ and a weaker one between $W$
and $C$ and $W$ and $\Delta$. The analyses presented in the preceding subsections have left open a reliable identification of their progenitors.
They found plausible that they include members of the Table \ref{tab4} sample, with $W$$>$0.3. But they could not exclude that they would also include members of the Table \ref{tab3} sample, with $W$$<$0.3. In particular, transition from Table \ref{tab3} to Table \ref{tab5} might occur via a regime of unstable pulsation. Both $W$$=$0 and $W$$=$1 curves are curves of constant luminosity. In the case of the former, the light spikes have become so narrow that they have disappeared, in spite of possibly having a large amplitude. In the case of the latter, the dips have become so narrow that they have disappeared, no matter how deep they were. A curve of constant luminosity is no longer variable and has lost memory of what it was before: if it becomes variable again, it may do it with low amplitude ($A$ small, $W$ irrelevant) or with short spikes of light ($W$ close to 0, $A$ irrelevant) or with short dips ($W$ close to 1, $A$ irrelevant).

\begin{figure*}
  \centering
  \includegraphics[width=18cm,trim={0cm 0 0cm 0},clip]{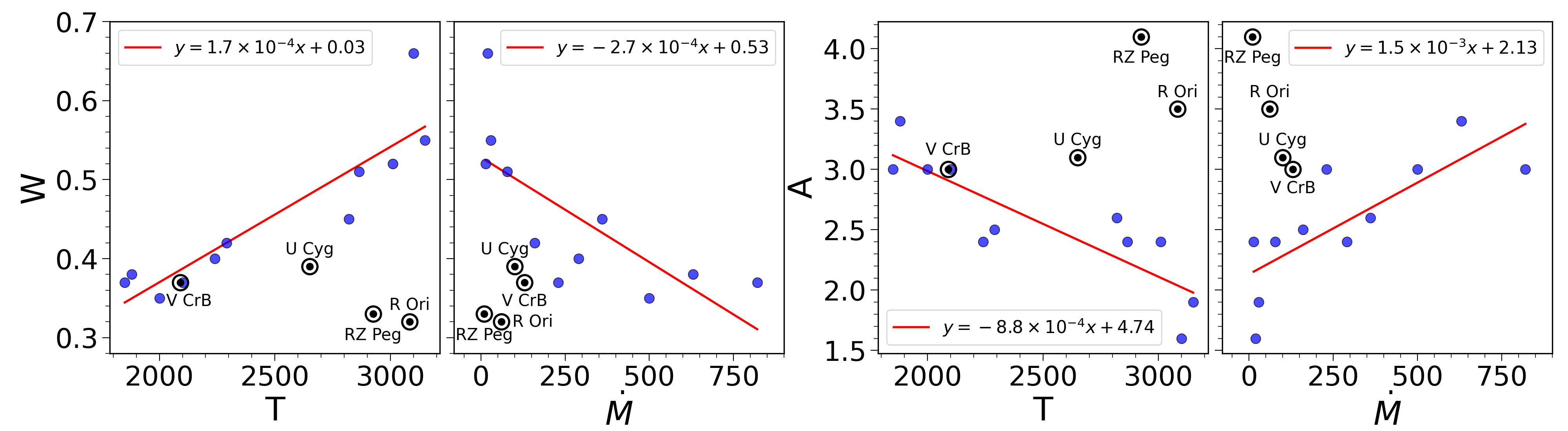}
  \caption{Dependence of the light curve parameters $W$ and $A$ on the effective temperature $T$(K) and mass loss rate $\dot{M}$ (10$^{-8}$ M$_\odot$ yr$^{-1}$)
    quoted by \citet{Bergeat2005} for 15 carbon-rich stars of the Table \ref{tab5} (C-rich) sample. Open circles show four stars
close to the Table \ref{tab4} ($W$$>$0.3) sample, V CrB, U Cyg, RZ Peg and R Ori, which are excluded from the linear fits displayed as red lines.}
 \label{fig5}
\end{figure*}

\citet{Bergeat2005} have collected data related to a large sample of carbon-rich stars, of which 15 are part of the
Table \ref{tab5} sample. Figure \ref{fig5} displays the dependence of the parameters $W$ and $A$ of their light curves on effective temperature $T$ and mass
loss rate $\dot{M}$. If one excludes four low $W$ curves that are close to the domain covered by the Table \ref{tab4} sample ($W$$>$0.3), V CrB, U Cyg, RZ Peg
and R Ori, $W$ tends to increase with temperature and decrease with mass loss rate while $A$ displays the opposite trend. Such a
behaviour would therefore suggest that the stars having large $W$ light curves evolve in the direction of decreasing $W$ when becoming
cooler. This trend is confirmed by the dependence of $W$ on the spectral index $i$ for spectral types C$i$: for $i$$<$6, namely warmer stars,
$<$$W$$>$$=$0.53$\pm$0.07 while for $i$$>$6, namely for cooler stars, $<$$W$$>$$=$0.39$\pm$0.04. This is however excluding R For, a clear outlier having a C4
spectral type and $W$$=$0.35.

\begin{figure*}
  \centering
  \includegraphics[width=18cm,trim={0cm 0 0cm 0},clip]{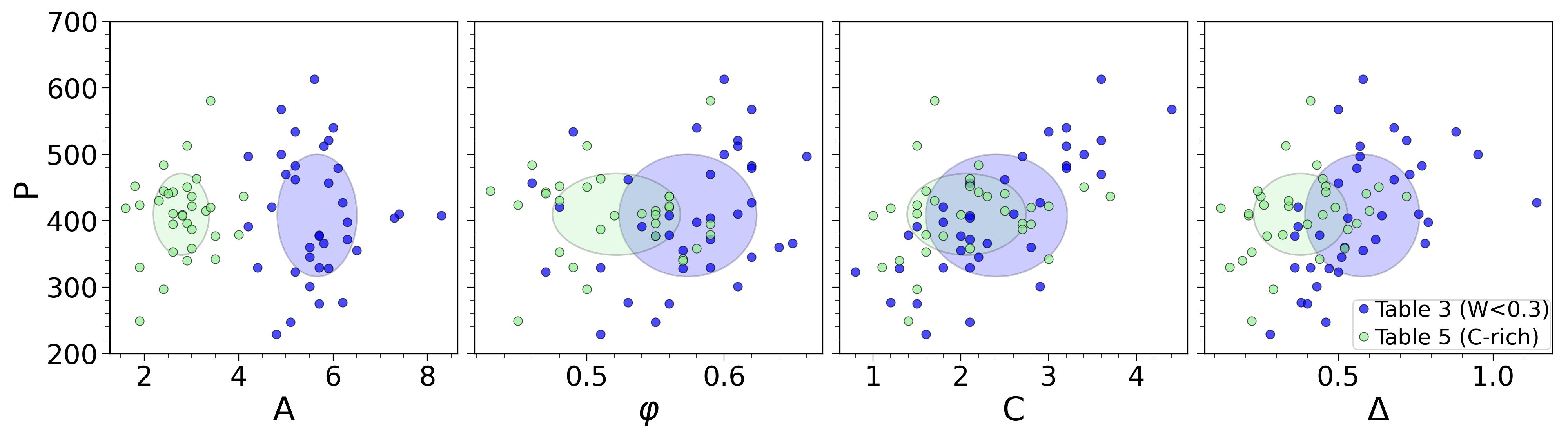}
  \caption{From left to right: projections of the parameter space distributions of the Table \ref{tab3} (blue) and Table \ref{tab5} (green) samples on the $P$
vs $A$, $P$ vs $\varphi_{\rm{min}}$, $P$ vs $C$ and $P$ vs $\Delta$ planes. Ellipses are centred on the mean coordinates and have half axes equal to their rms deviations.}
 \label{fig6}
\end{figure*}

To the extent that we can take as granted that carbon stars keep cooling down \citep{Marigo2002}, this suggests that the carbon-rich stars of
the Table \ref{tab5} sample having curves of large $W$ value enter it from the large $W$ side and evolve toward smaller values of $W$. If such is the
case, they would probably have a Table \ref{tab3} ($W$$<$0.3) origin: the larger $W$, the smaller the oscillation amplitude (Figure \ref{fig4}), namely the closer
from a regime of unstable pulsation. Figure \ref{fig6} illustrates what such a transition would look like by displaying projections on the $P$ vs $A$,
$P$ vs $\varphi_{\rm{min}}$, $P$ vs $C$ and $P$ vs $\Delta$ planes of the distributions in parameter space of the Table \ref{tab3} and Table \ref{tab5} samples. It shows that, apart from
$A$ and $W$, a transition from Table \ref{tab3} to Table \ref{tab5} is reasonably continuous in the other parameters. The discontinuity in amplitude is not
surprising: absorption of the dust envelope is known to be larger for carbon-rich rather than oxygen-rich stars. One may instead assume that Table \ref{tab3} stars such as $\chi$ Cyg will become irregular carbon-rich stars absent from catalogues of Mira variables. This would leave the identity of the precursors of the large $W$ members of the Table \ref{tab5} curves unanswered, but one might similarly assume that they are irregular curves of S spectral type stars. In this context, one might similarly argue that the low colour index stars of the Table \ref{tab4} sample, with S spectral types, would also evolve toward irregular carbon-rich stars. These comments illustrate the difficulty to reliably interpret what is observed.

RS Cyg and V Oph, both parts of the Figure \ref{fig5} sample, have large values of $W$, 0.66 and 0.52, respectively. They would then
be likely successors of the Table \ref{tab3} sample. RS Cyg has a high temperature of 3100 K and a low mass loss rate of 2$\times$10$^{-7}$ M$_\odot$yr$^{-1}$
 \citep{Guandalini2006}. Its light curve has been extensively studied and commented by \citet{Cadmus2022, Cadmus2024}. It often
displays a hump on the ascending branch, close to maximum light, possibly producing a double-maximum profile. V Oph has been the
target of detailed VLT observations in the infrared \citep{Ohnaka2007, Rau2019, Hulberg2025} suggesting an effective
temperature of $\sim$2700 K while Figure 5 quotes a temperature of 3010 K and a mass loss rate of 1.4$\times$10$^{-7}$ M$_\odot$yr$^{-1}$.
However, the identification of the origin of curves having lower values of $W$ turns out to be difficult. BH Cru and LX Cyg
were shown by \citet{Uttenthaler2011, Uttenthaler2016} to have recently switched from spectral type S to C: one would naively expect
to find them near an edge of the domain covered by the carbon-rich sample in the parameter space; but such is not the case, their curves have
values of $W$ (0.47 and 0.46, respectively) that sit in the middle of those of the Table \ref{tab5} sample, a bit on the high side: they may be
successors of the Table \ref{tab3} ($W$$<$0.3) sample as well as of the Table \ref{tab4} ($W$$>$0.3) sample, although the value of the temperature \citep[$\sim$3000 K,][]{Zijlstra2004}
rather favours the former.

\citet{Rau2017} and \citet{Paladini2017} have published a detailed comparison of the infrared spectra of R Lep and R Vol
obtained by VLTI/MIDI. As the former has a $W$ value of 0.42 while the latter has the smallest $W$ value in the sample (0.31), one might
expect such a comparison to reveal significant differences between the two stars. But it does not: the analyses presented by \citet{Rau2017}
produce instead very similar stellar parameters: a luminosity of 8000-8500 solar luminosities and an effective temperature of
$\sim$2900 K. R Lep is part of the Figure \ref{fig5} sample with a mass loss rate of 1.6$\times$10$^{-6}$ M$_\odot$yr$^{-1}$. Both R Lep and R Vol are close
to RZ Peg, R Ori and U Cyg, which have similar temperatures and low $W$ values.

\begin{deluxetable*}{lcl ccc ccc c}%{rrrrrrrrrrr}%
\tablenum{5}
\tablecaption{List of curves of C spectral type.\label{tab5}}
%\tablewidth{0pt}
\tablehead{
  \colhead{Name}&\colhead{$P$ (days)}&\colhead{Spctr}&\colhead{Tc}&\colhead{$^{12/13}$C}&\colhead{$C$ (mag)}&\colhead{$A$ (mag)}&\colhead{$W$}&\colhead{$\Delta$ (mag)}&\colhead{$\varphi_{\rm{min}}$}}
\startdata
AZ Aur	&415	&C8 	&yes	&$-$	&2.5	&3.3	&0.37	&0.6	&0.55	\\
R Cap	&342	&C?	&$-$	&$-$	&3	&3.5	&0.35	&0.44	&0.57	\\
AX Cep	&396	&C?	&$-$	&$-$	&2.7	&2.9	&0.39	&0.56	&0.55	\\
S Cep	&484	&C7	&yes	&224	&2.2	&2.4	&0.4	&0.43	&0.46	\\
V CrB	&358	&C6	&yes	&$-$	&2.1	&3	&0.37	&0.52	&0.58	\\
U Cyg	&463	&C9 	&yes	&14	&2.1	&3.1	&0.39	&0.45	&0.51	\\
V Cyg	&420	&C7 	&$-$	&20	&2.8	&3.4	&0.38	&0.49	&0.56	\\
T Dra	&422	&C6 	&yes	&24	&3	&3	&0.37	&0.34	&0.56	\\
R For	&387	&C4 	&$-$	&$-$	&2.7	&3	&0.35	&0.53	&0.51	\\
T Lyn	&409	&C7 	&$-$	&$-$	&2	&2.8	&0.44	&0.45	&0.55	\\
R Ori	&377	&C8	&$-$	&$-$	&1.8	&3.5	&0.32	&0.27	&0.55	\\
BH Cru	&513	&C8 	&yes	&8	&1.5	&2.9	&0.47	&0.33	&0.5	\\
LX Cyg	&581	&C?	&yes	&32	&1.7	&3.4	&0.46	&0.41	&0.59	\\
W Cas	&408	&C7	&yes	&25	&1	&2.8	&0.45	&0.21	&0.52	\\
UV Aur	&395	&C8 	&yes	&$-$	&2.8	&2.6	&0.43	&0.4	&0.59	\\
V Aur	&353	&C6	&$-$	&$-$	&1.6	&2.6	&0.45	&0.22	&0.48	\\
RZ Peg	&437	&C9	&yes	&9	&2.3	&4.1	&0.33	&0.25	&0.56	\\
RU Vir	&437	&C8	&yes	&$-$	&3.7	&3	&0.37	&0.72	&0.56	\\
R Vol	&451	&C?	&$-$	&$-$	&3.4	&2.9	&0.31	&0.63	&0.5	\\
VX Gem	&379	&C9 	&yes	&9	&1.6	&4	&0.36	&0.32	&0.59	\\
S Cam	&330	&C6 	&$-$	&14	&1.1	&1.9	&0.56	&0.15	&0.49	\\
X Cas	&424	&C5	&$-$	&19	&1.5	&1.9	&0.52	&0.26	&0.45	\\
RV Cen	&445	&C3	&$-$	&$-$	&1.6	&2.4	&0.51	&0.24	&0.43	\\
RS Cyg	&419	&C5.5	&$-$	&$-$	&1.2	&1.6	&0.66	&0.12	&$-$	\\
WX Cyg	&411	&C6 	&yes	&5	&1.5	&2.6	&0.52	&0.21	&0.54	\\
U Lyr	&452	&C4	&$-$	&23	&2.1	&1.8	&0.54	&0.46	&0.48	\\
V Oph	&297	&C4 	&yes	&11	&1.5	&2.4	&0.52	&0.29	&0.5	\\
Y Per	&249	&C5 	&$-$	&$-$	&1.4	&1.9	&0.55	&0.22	&0.45	\\
PQ Cep	&443	&C6	&$-$	&$-$	&2.2	&2.6	&0.46	&0.46	&0.47	\\
R Lep	&441	&C7	&$-$	&34	&2.5	&2.5	&0.42	&0.59	&0.47	\\
BD Vul	&430	&C6	&$-$	&$-$	&1.7	&2.3	&0.45	&0.27	&0.48	\\
R CMi	&340	&C0	&yes	&11	&1.3	&2.9	&0.47	&0.19	&0.57	\\
\enddata
\end{deluxetable*}

S Cep, with $W$$=$0.40, has an effective temperature of 2100 K \citep{Nowotny2005, Nowotny2010}, leaving open both a Table \ref{tab3} ($W$$<$0.3) and a
Table \ref{tab4} ($W$$>$0.3) origin. Such is indeed the case for all curves having $W$ values of this order or smaller. The 13 curves having $W$$<$0.4 cover
rather uniformly the parameter space, failing to reveal a clear separation between two groups of curves. These examples illustrate the
difficulty of producing reliable interpretations of the evolution of the carbon-rich stars having curves in the Table \ref{tab5} sample:
  the likely coexistence of Table \ref{tab3} ($W$$<$0.3) and Table \ref{tab4} ($W$$>$0.3) successors is a major complication and a reliable distinction between them is difficult.

Another source of complication is the large fraction of carbon-rich stars in an unstable regime of pulsation, implying that
their light curves are often excluded from our sample and cannot be parameterized reliably in the form used in the present article.
\citet{Marigo2020} have produced a very insightful description of the evolution of carbon-rich stars along the AGB, paying attention
to the initial mass of the star, the TDU efficiency, the C/O ratio and other quantities of utmost relevance. In particular, they have
shown the presence of a kink in the initial-to-final mass relation, occurring at initial masses just below two solar masses. They suggest
(see their Figure 6 in Supplementary information) that its progenitors spend a fraction of their carbon star life as semi-regular
variables, characterised by small amplitude pulsations, low C/O ratio, small mass-loss rates and low terminal velocities. A detailed
study, case by case, of the curves of Table \ref{tab5} should be made within the framework of their study; however, this is well beyond the
scope of the present article.

\begin{figure*}
  \centering
   \includegraphics[width=18cm,trim={0cm 0 0cm 0},clip]{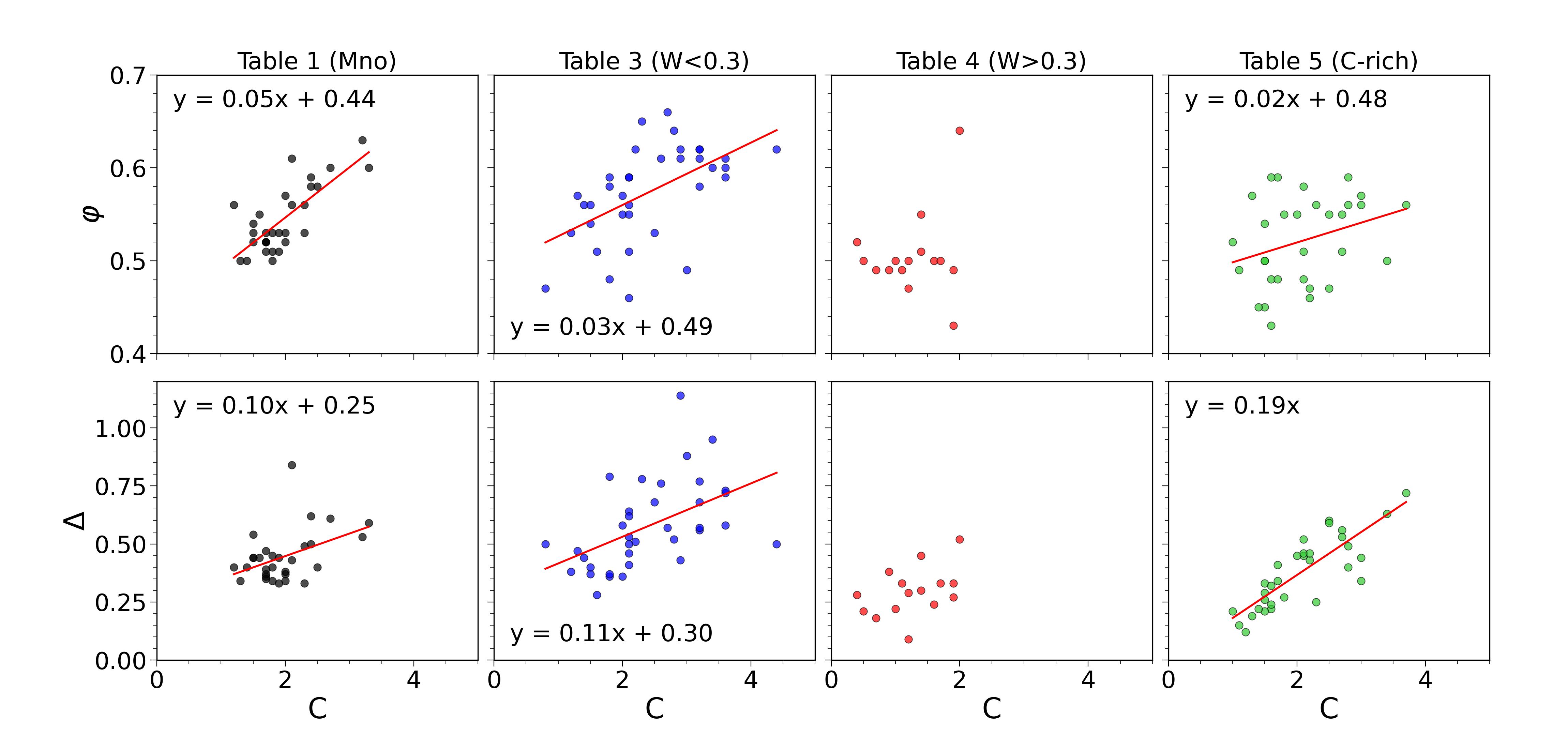}
  \caption{Dependence of $\varphi_{\rm{min}}$ (upper row) and $\Delta$ (lower row) on $C$ for the samples of Tables \ref{tab1} (Mno), \ref{tab3} ($W$$<$0.3), \ref{tab4} ($W$$>$0.3) and \ref{tab5} (C-rich), respectively from left to right. Linear fits are shown for the Table \ref{tab1}, \ref{tab3} and \ref{tab5} samples.}
 \label{fig7}
\end{figure*}

Finally, we recall that we have shown in Paper I (Section 5, Figure 14) the dependence of $\Delta$ and $\varphi_{\rm{min}}$ on $C$ and remarked that carbon-rich stars and
oxygen-rich stars follow similar trends, the former having slightly lower values of $\Delta$ and of $\varphi_{\rm{min}}$ than the latter for a same value of $C$.
Figure \ref{fig7} illustrates this issue in the present sample, displaying separately the samples of Tables \ref{tab1}, \ref{tab3}, \ref{tab4} and \ref{tab5}. The Table \ref{tab1} (Mno) distributions
illustrate the transition from what was defined as first and second families of Mno curves in Paper I, with both $C$ and $\varphi_{\rm{min}}$ increasing;
the location of the curve of R Cas ($\Delta$$=$0.84) confirms the suggestion that was made in Subsection 3.2 that the star is in the process of
switching to the Myes spectral type. The same trend is observed in the distributions of the Table \ref{tab3} ($W$$<$0.3) sample, with a global shift toward larger $C$ and $\varphi_{\rm{min}}$; a clear shift toward large values of $\Delta$ gives evidence for the increased instability of the pulsation regime. The
concentration in small values of both $\Delta$ and $C$ illustrates the difference between the curves of the Table \ref{tab4} ($W$$>$0.3) sample and the others. The
distributions displayed by the Table \ref{tab5} sample confirm the comments made in Paper I.

\subsection{Humpy curves and hump-less curves}
\begin{figure*}
  \centering
  \includegraphics[width=18cm,trim={0cm 0 0cm 0},clip]{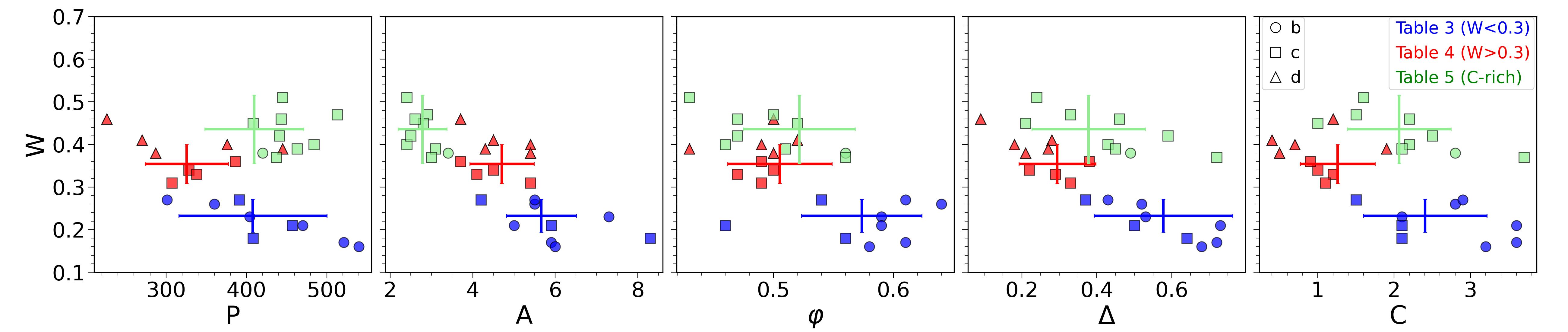}
  \caption{Distribution of the humpy curves in the parameter space. Different colours are used to distinguish between the Table \ref{tab3} ($W$$<$0.3), \ref{tab4} ($W$$>$0.3) and \ref{tab5} (C-rich) samples and different symbols to distinguish between hump types \textbf{\textit{b}}, \textbf{\textit{c}} and \textbf{\textit{d}} as shown in the inserts.}
 \label{fig8}
\end{figure*}

In Paper I, we remarked that when a hump was observed on the ascending branch, there was a relation between its location and the
parameters $C$, $\Delta$ and $\varphi_{\rm{min}}$: namely the closer it was to minimal light, the larger were the values of these three parameters. Figure \ref{fig8}
illustrates the present situation. We selected curves displaying a clear hump on the ascending branch and classified them as type \textbf{\textit{b}}, \textbf{\textit{c}} or
\textbf{\textit{d}} according to the hump location: \textbf{\textit{b}} for humps close to minimal light, \textbf{\textit{d}} for humps close to maximal light and \textbf{\textit{c}} in between. In the
definitions of both what we call a strong hump and of which class the curve belongs to, there is of course some subjectivity which
must be kept in mind. Moreover, what we define as humps are rather abrupt changes of the rate of increase of the luminosity:
while \textbf{\textit{c}} types are real humps, \textbf{\textit{b}} and \textbf{\textit{d}} types are rather  flattening of the light minimum and maximum, respectively.
Figure \ref{fig8} reveals a number of remarkable features: all \textbf{\textit{d}} types are from the Table \ref{tab4} sample and, with the
exception of V Cyg, all \textbf{\textit{b}} types are from the Table \ref{tab3} sample; instead, \textbf{\textit{c}} types are present in all three samples and span a very large
range of $W$ values; the general trends observed in Paper I are clearly confirmed and complemented with the dependence of the hump
type on $P$, $A$ and $W$. Such humps may simply be the trivial effect of non-linear dynamics, similar humps are observed in RR Lyrae
stars and in bump Cepheids with the Hertzsprung progression. In such a case, they may not be amenable to a simple interpretation,
such as tracking a shock wave as suggested by \citet{Kudashkina1994}. Yet, their distribution in the parameter space suggests
that they carry some additional information, which we however fail to decrypt.

\section{Other stars}
\subsection{Hot Bottom Burners}
\begin{figure*}
  \centering
 \includegraphics[width=3.33cm,trim={0cm -0.3cm 0cm 0},clip]{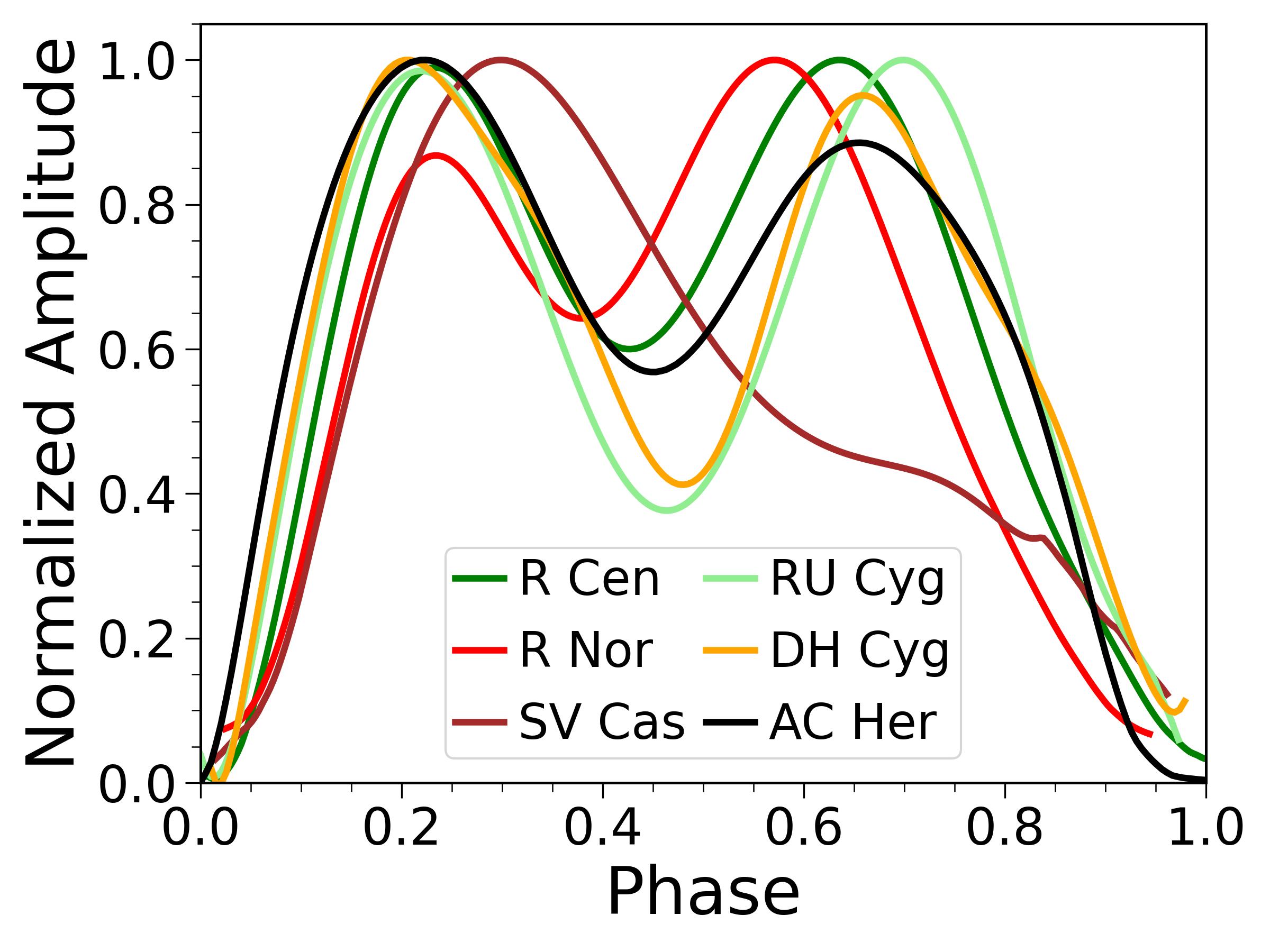}
\includegraphics[width=3.33cm,trim={0cm -0.3cm 0cm 0},clip]{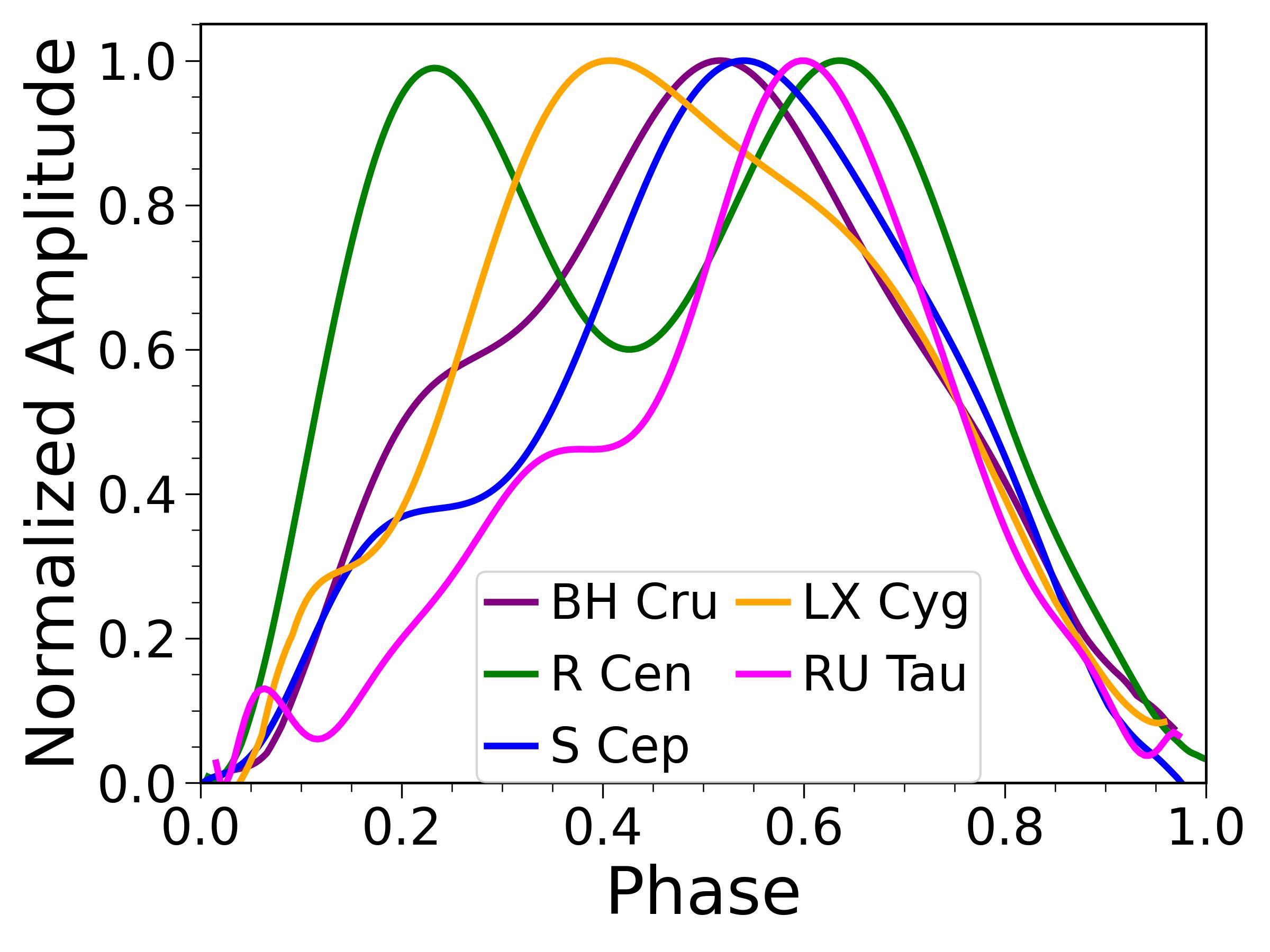}
\includegraphics[width=3.4cm,trim={0cm 0 1.5cm 0},clip]{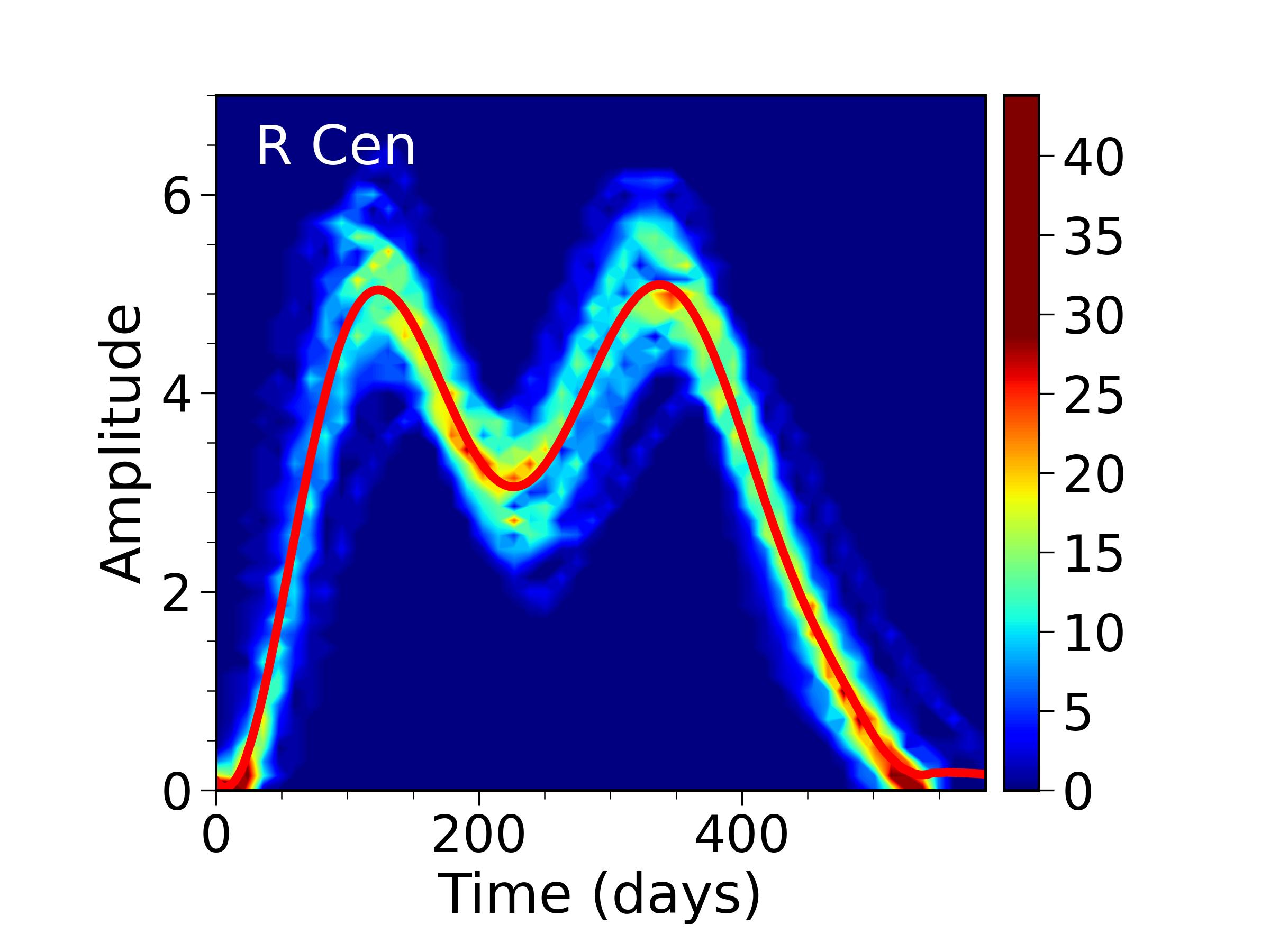}
\includegraphics[width=3.4cm,trim={0cm 0 1.5cm 0},clip]{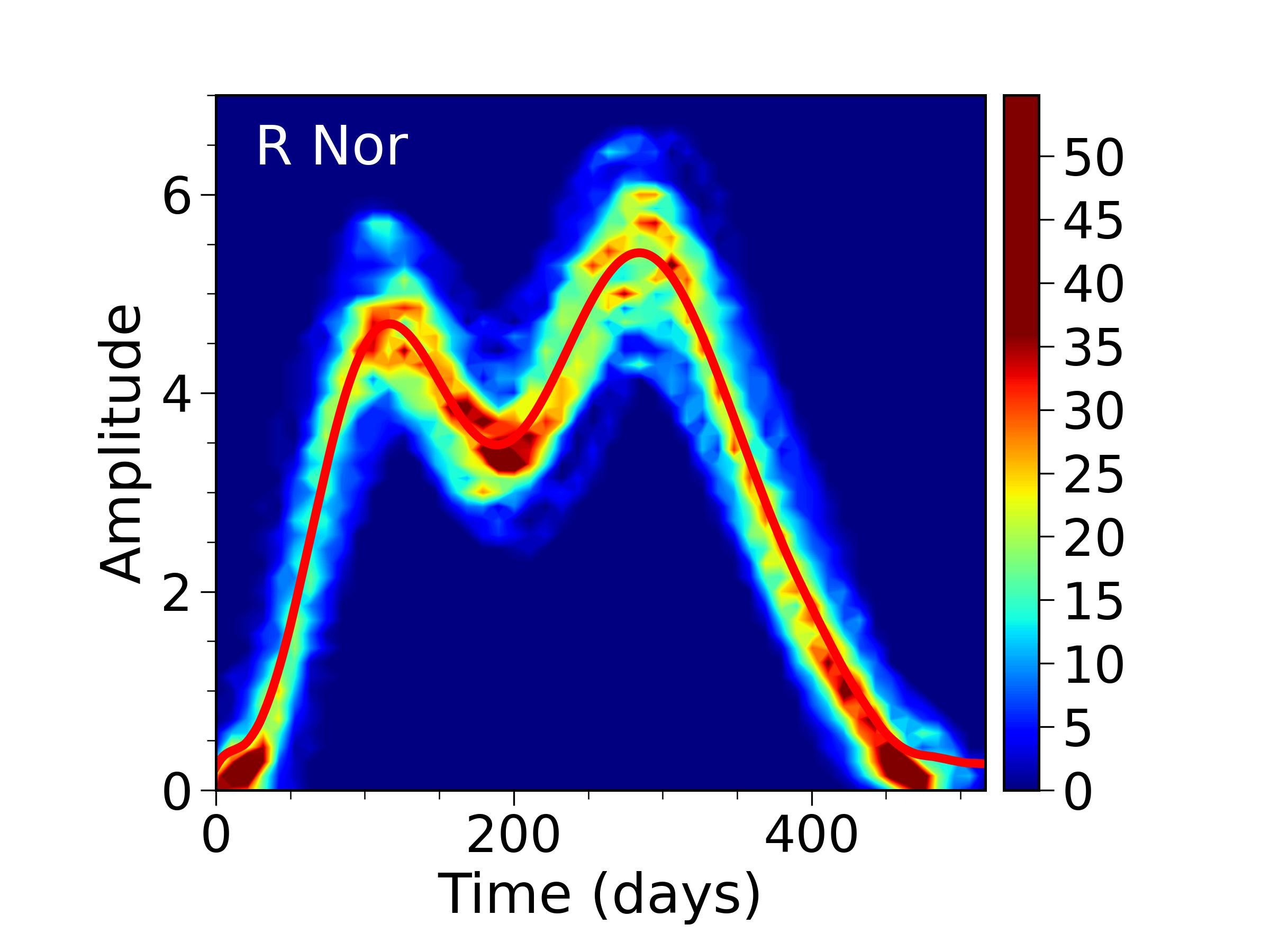}
\includegraphics[width=3.4cm,trim={0cm 0 1.5cm 0},clip]{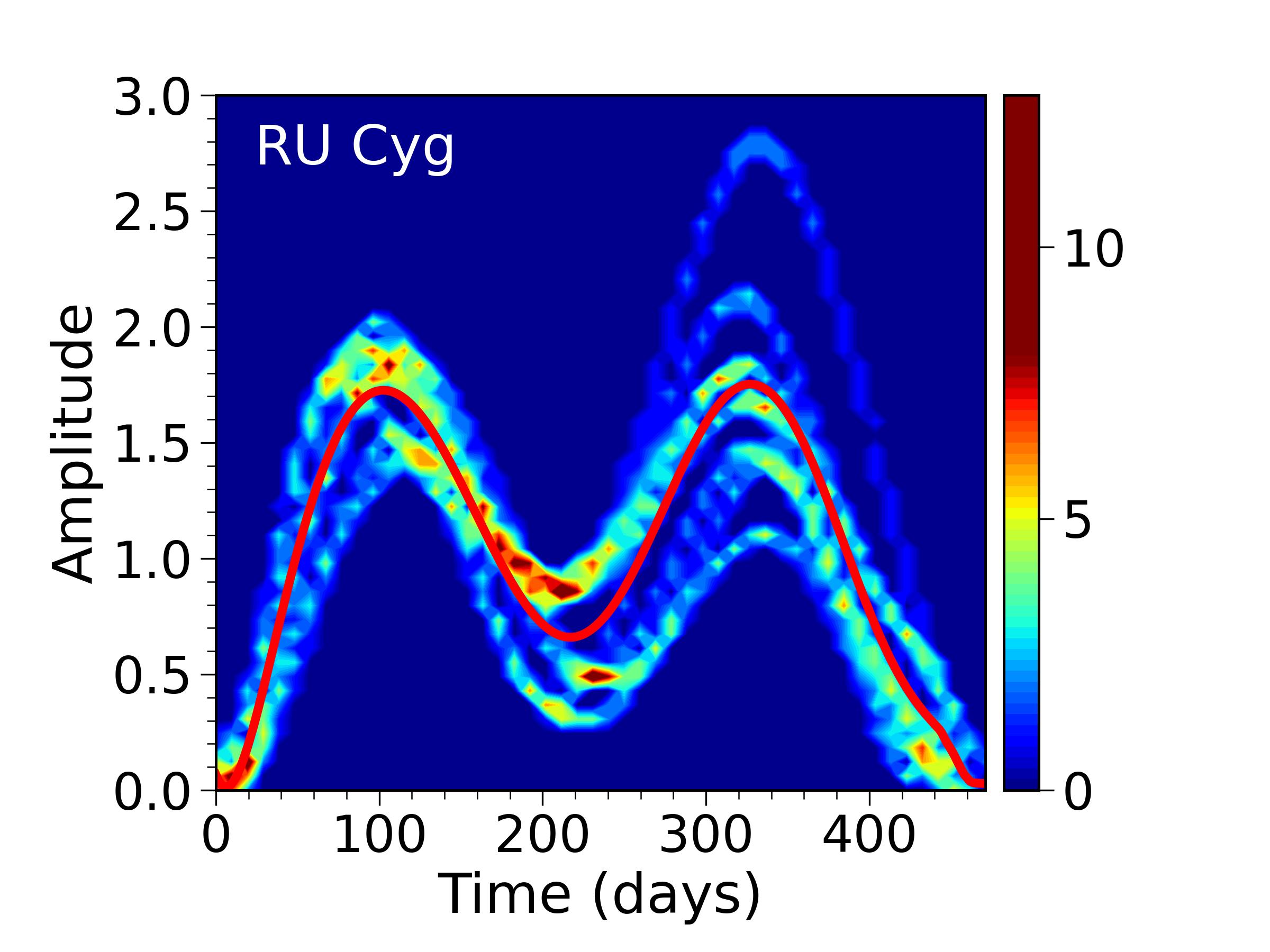}
\includegraphics[width=3.4cm,trim={0cm 0 1.5cm 0},clip]{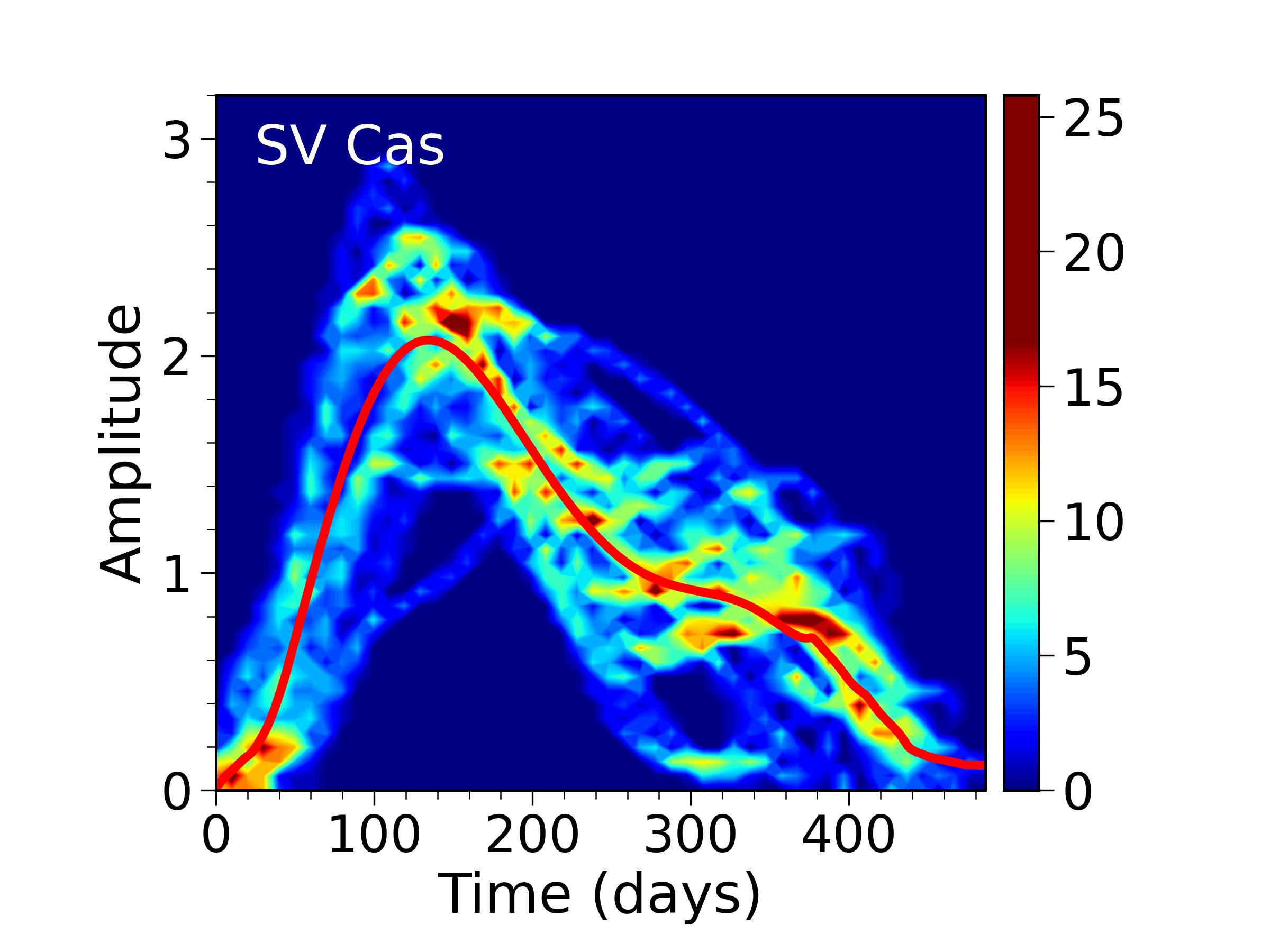}
\includegraphics[width=3.4cm,trim={0cm 0 1.5cm 0},clip]{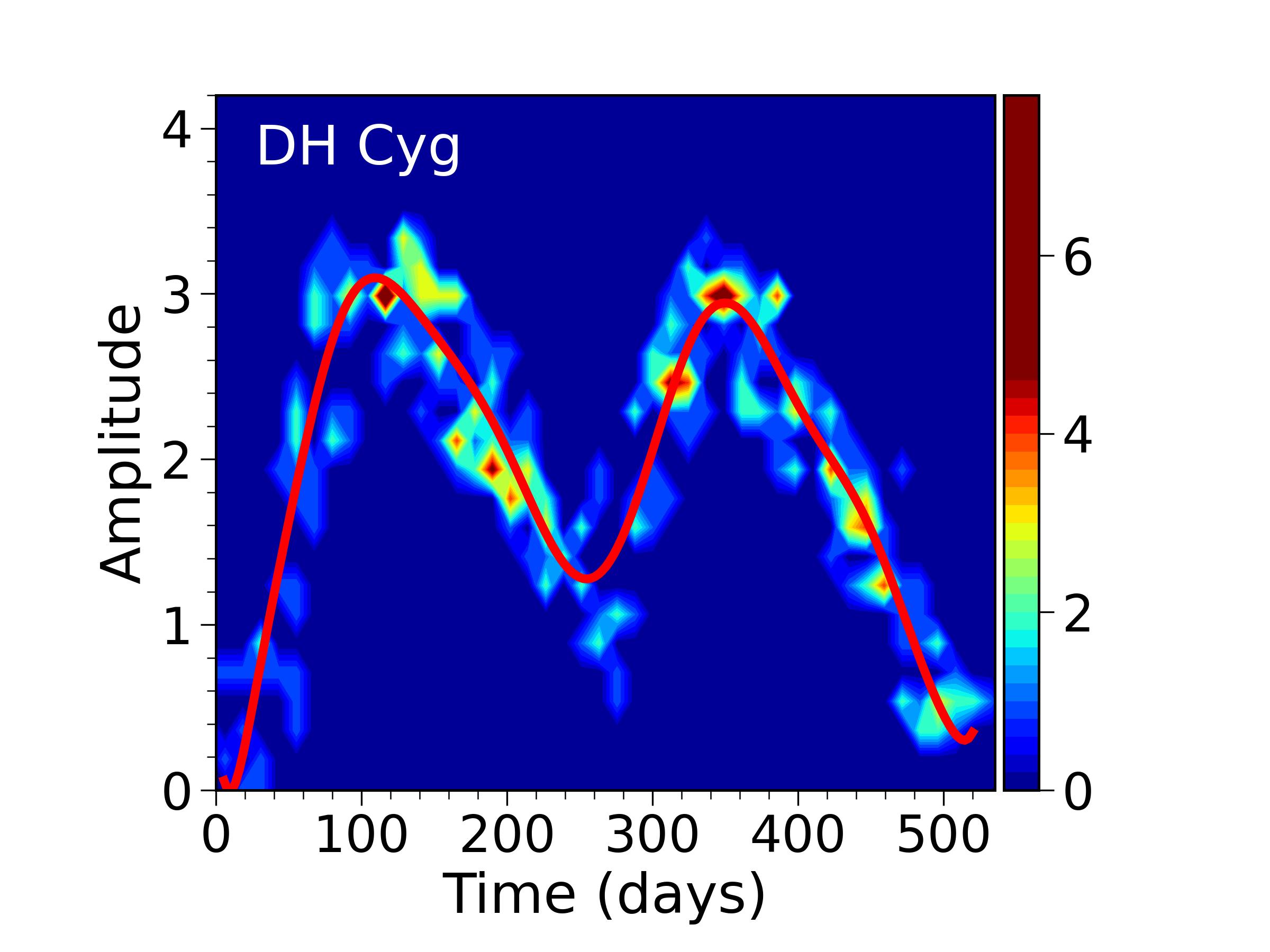}
\includegraphics[width=3.4cm,trim={0cm 0 1.5cm 0},clip]{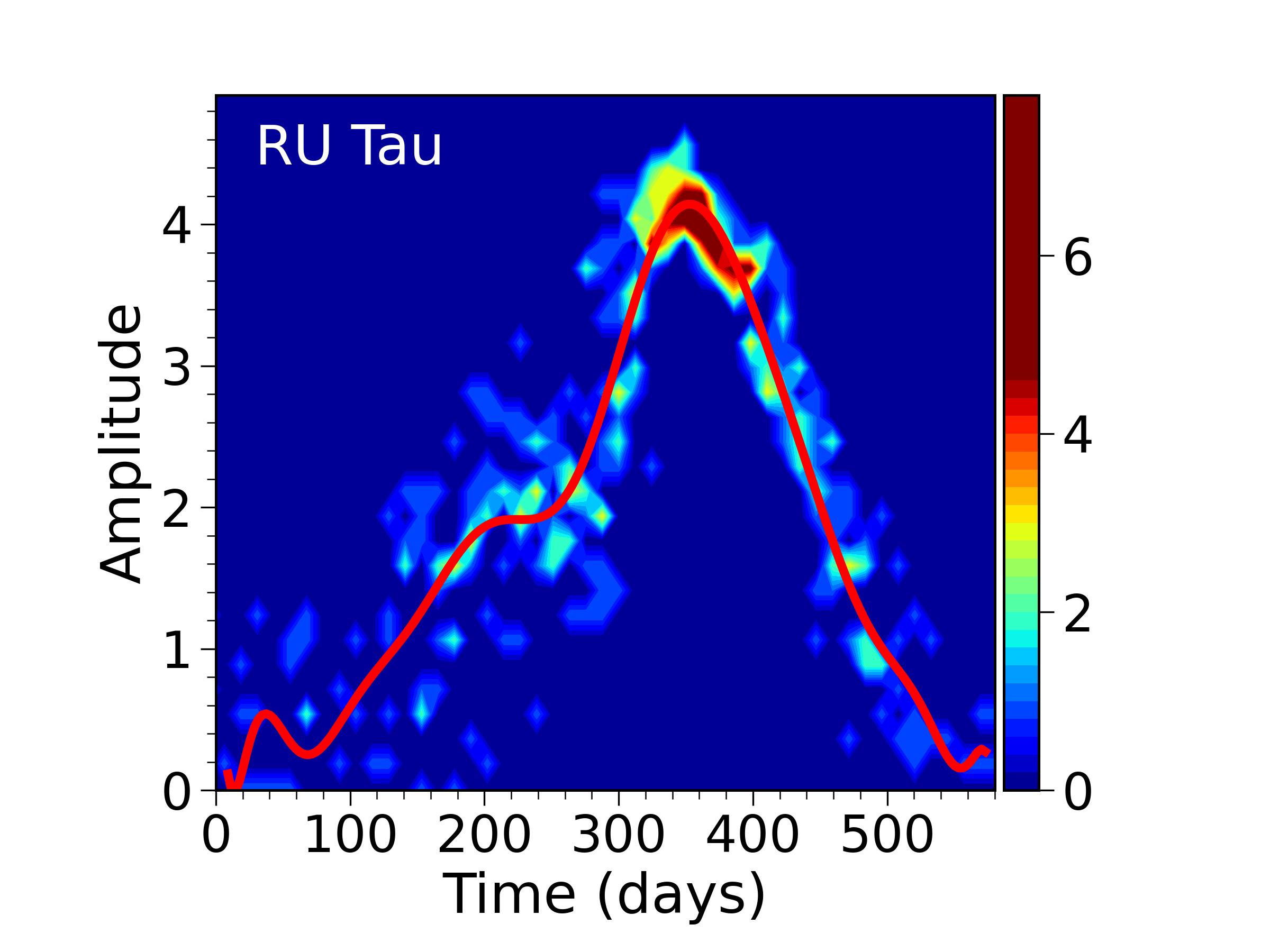}
\includegraphics[width=3.4cm,trim={0cm 0 1.5cm 0},clip]{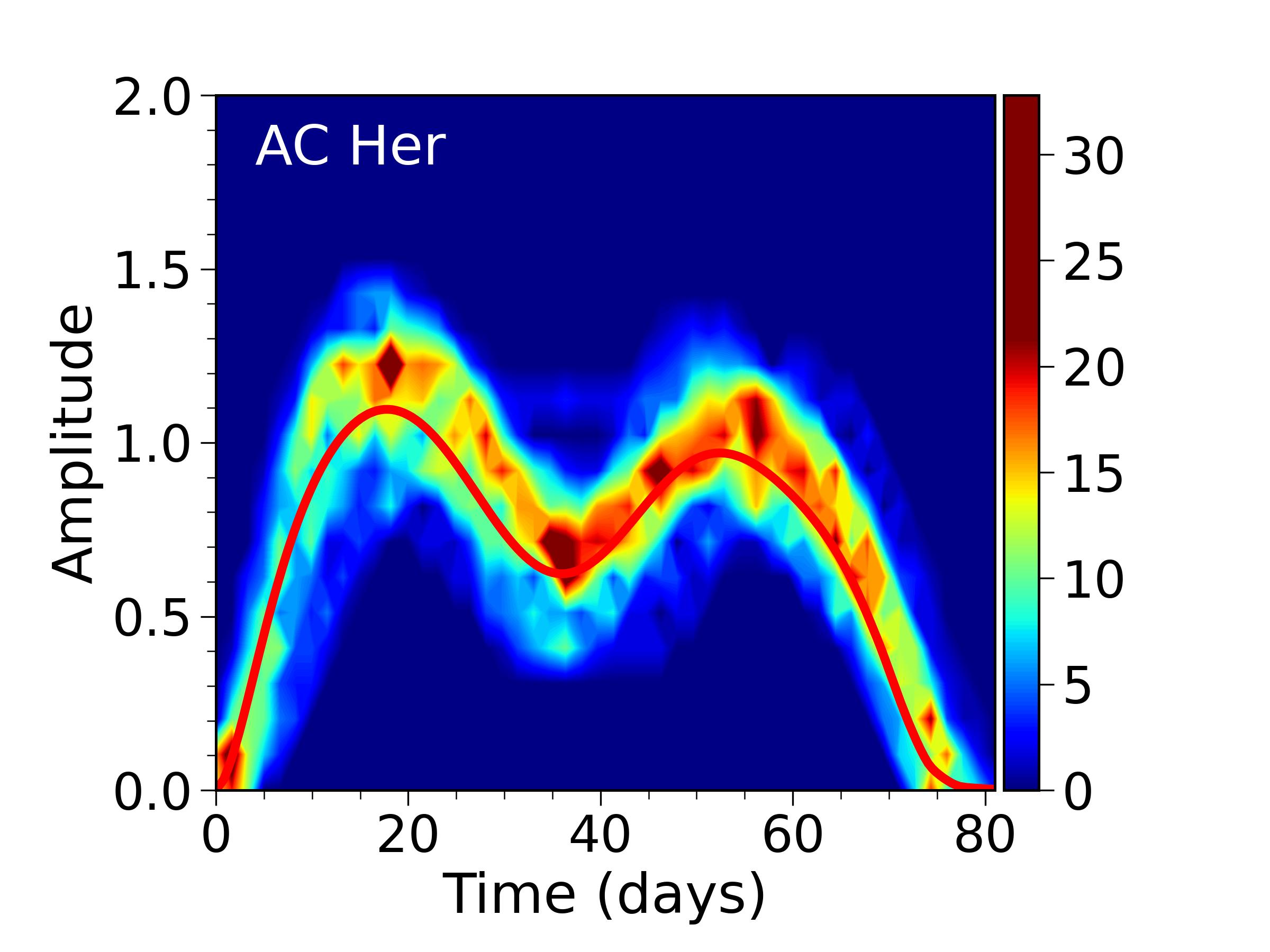}

  \caption{The pair of left panels compares the normalised profiles of the light curves of different stars with that of R Cen; the leftmost
panel includes curves of similar shapes, the second panel includes curves of different shapes (see text). The following panels display
superimposed profiles of cycles taken from the light curves of stars discussed in the text: the four well-established HBB candidates, R
Cen, R Nor, RU Cyg and SV Cas, the possible additional HBB candidate, DH Cyg, the particular case of RU Tau, and the archetypal
RV Tau star, AC Her.}
 \label{fig9}
\end{figure*}

R Cen and R Nor have been identified as clear Hot Bottom Burning (HBB) candidates at the beginning of the TP-AGB by \citet{GarciaHernandez2013}
and \citet{Uttenthaler2011,Uttenthaler2012} on the basis of detailed evaluations of the abundances of technetium,
lithium, rubidium and zirconium in the spectra: they are Tc-poor and Li-rich and have low mass-loss rates at the level of 10$^{-7}$ M$_\odot$ yr$^{-1}$,
properties which they share with two other clear HBB candidates, SV Cas and RU Cyg, both semi-regular variables.
They are expected to have initial masses of 4 to 6 solar masses, in contrast to more massive HBB candidates with very high mass-loss
rates ($\sim$10$^{-5}$ M$_\odot$ yr$^{-1}$), known as OH/IR stars, at the end of the TP-AGB \citep{GarciaHernandez2006, GarciaHernandez2007, GarciaHernandez2009}.
Their light curves are double-peaked (Figure \ref{fig9}) and that of R Cen has recently experienced a sudden change in period and
amplitude. Moreover, they are spectacular outliers in colour vs period diagrams \citep{Uttenthaler2019, MerchanBenitez2023,Whitelock2000}.
They have been commented upon in numerous earlier studies, both for R Cen \citep{Keenan1966,Hawkins2001, Walker2001, Templeton2005, Garcia2006, Vogt2016, Rea2024}
and for R Nor \citep{Vogt2016}. Most authors take it as granted that the sudden changes observed in the light curve of R Cen
are the result of a recent thermal pulse, as had been first proposed by \citet{Wood1981}. Measurements of the radial velocity in the
near infrared \citep{Lebzelter2005} have given evidence for a clear anti-correlation with the light curve.

The light curve of SV Cas, rather than being double-peaked, displays a hump on the descending branch, which is exceptional.
The amplitude of the oscillation increased from $\sim$1.5 mag at the end of the 1980s to $\sim$ 3 mag today. The mass loss rate has been
measured as 2.4$\times$10$^{-7}$ M$_\odot$ yr$^{-1}$ \citep{Scicluna2022} and the $^{12}$C/$^{13}$C ratio as 10$\pm$2 \citep{Hinkle2016}. Over the past
century, the light curve of RU Cyg \citep{Cadmus2021, Cadmus2024} has evolved from a relatively regular sequence of double-peaked cycles to
more irregular cycle profiles; while the period has stayed at its value of $\sim$468 days, the amplitude of the oscillations has decreased
from nearly 2 to $\sim$0.5 units of magnitude. Both SV Cas and RU Cyg light curves (Figure \ref{fig9}) share with those of R Cen and R Nor the
relative brevity of the ascending and descending branches, implying a large value of $W_{\rm{1/5}}$. We noted that another star displays a light
curve that resembles those of the four clearly identified HBB candidates: DH Cyg (Figure \ref{fig9}). The star has been studied by \citet{Smith2014}
as a metal-poor field giant and its emission in the mid-infrared has been found to be very strong, together with AC Her.
The recent changes experienced by the R Cen curve and the changing regime of pulsation displayed by the curves of SV Cas and RU
Cyg are evidence for HBB stars at the beginning of the TP-AGB to be close to the limits of the domain of stable pulsation.

The double-peaked profiles of R Cen and R Nor are outstanding. The $W_{\rm{R}}$ and $W_{\rm{1/5}}$ parameters do not obey the standard
relation between them, the former being not as large as implied by the latter because of the depression between the two maxima. We
looked for possible other cases among Long Period Variables but did not find any convincing case. An example is shown in the
second panel of Figure \ref{fig9} that compares mean profiles of curves sharing with R Cen and R Nor the property of having both a large
value of the width parameter, $W$, and a small value of the colour index, $C$. The curves of BH Cru, LX Cyg and S Cep, carbon-rich stars
of which the light curves are occasionally quoted as double-peaked, as are those of TT Cen, RS Cyg and U CMi, are of a different
nature: they simply display a hump on the ascending branch. So does the curve of RU Tau, a Mira variable of spectral type M3.5
quoted by \citet{MerchanBenitez2023} as having a very large meandering period of 589 days and a colour index $C$$=$1.9; it is the only
oxygen-rich Mira variable sharing with R Cen and R Nor a large period, low colour indices and a large width ($W$$=$0.41), and as such
deserves being given careful attention; however, the profile of its light curve favours a Table \ref{tab4} rather than HBB assignment.

In contrast with the light curves of AGB stars, those of RV Tauri stars commonly display double-peaked cycles of a shape
similar to those displayed by R Cen and R Nor (Figure \ref{fig9}). RV Tauri stars \citep[][and references therein]{Manick2019, Gezer2015}
are thought to be metal-deficient, population II, low mass post-AGB stars, often surrounded by a dust disc, generally
associated with the presence of a Main Sequence companion, which causes periodic obscuration of the light curve. Their periods range
between $\sim$30 and $\sim$150 days and, having nearly lost their circumstellar envelope, they have low masses, typically between 0.3 and 0.9
solar masses. Numerous detailed studies of the shapes of the light curves \citep{Tuchman1993, Fokin1994, Buchler1995, Kollath1998}
have shown that they display a broad range of irregularity. Whether the understanding of the dynamics governing
the pulsations of an archetypal RV Tau star, AC Her (Figure \ref{fig9}), may help or not with the understanding of the dynamics governing the
pulsations of R Cen and R Nor is unclear. On the one hand, the similarity between the light curves is so striking that one feels that the
dynamics must share some common features. On the other hand, one may fear that such similarity is the simple result of the basic
properties of non-linear dynamics, which can be expressed in apparently simple terms using the language of fundamental and overtone
modes, but which fail to tell much about the actual physical mechanisms at stake \citep{Fokin1994}, which, in the cases of AC Her vs R
Cen/R Nor, are obviously very different.

\subsection{Mno outliers}
Five stars of Mno spectral type were excluded from the sample listed in Table \ref{tab1} as having light curves presenting peculiarities that
deserve a special discussion, W Dra, W Hya, T Ari, U CMi and T UMi.

The curve of W Dra is only marginally different from those of the Table \ref{tab1} sample. While it was reasonable to exclude it from
  the studied sample, it is now clear that it might as well have been included, which would have had essentially no impact on the conclusions
  of the analysis. We did not find in the published literature evidence for the star to display abnormal features.

The curve of W Hya is instead very different from those of the Table \ref{tab1} sample and closely resembles those of some stars of
spectral type Myes listed in Table \ref{tab4}, such as R Hya. \citet{Uttenthaler2011} give a detailed description of the search for technetium in
the spectra of R Hya and W Hya, both displaying significant period changes. They conclude positively for the former and negatively
for the latter but note that intrinsically weak lines may go entirely undetected if a Mira star is observed at a pulsation phase with severe
line weakening. Being nearby, W Hya has been the target of a very large number of high-resolution observations, each revealing
detailed features that could not have been detected if the star were more distant. Recent examples quoting relevant references are \citet{Hoai2022b} and \citet{Khouri2020}.
The curve of T Ari is similar to curves of typical carbon-rich stars such as WX Cyg. Together with that of U CMi, its profile
has a very large $W$ parameter, $\sim$0.5, which is otherwise found only for carbon-rich stars or Hot Bottom Burners. Unfortunately, too
little is known about these stars to suggest plausible explanations and the cases must remain unexplained.

Finally, T UMi is well-known as a case of sudden period change, with a decline from more than 300 days to about 200 days
over the past 20 years, causing it to leave the domain of stable pulsation. Before this change, the parameters of the light curve were as
listed in Table \ref{tab6}, indeed very different from those of Table \ref{tab1}. \citet{Mattei1995}, \citet{Gal1995} and \citet{Szatmary2003}
suggested that T UMi is currently just before the beginning of a helium shell flash; \citet{Uttenthaler2011} remarked that, in
such a case, the lack of Tc and Li in the spectrum imply that T UMi is either not massive enough to experience a TDU event, or that
the thermal pulse is early and/or weak. Indeed, both \citet{Fadeyev2022} and \cite{Molnar2019} have recently produced state-of-the-art
models consistent with such an interpretation, showing that T UMi had an initial mass between 1.0 and 1.5 solar masses (1.2$\pm$0.2 for
Moln\'{a}r et al.). This makes T UMi very different from the Hot Bottom Burning R Cen and illustrates how similarity of light curves
cannot be naively interpreted as similarity of underlying dynamics.

\subsection{Binaries}
$o$ Cet and R Aqr are well-known Mira variables of spectral type Myes, which were not retained in the selection of \citet{MerchanBenitez2023}
as being members of binaries. Their curve parameters are listed in Table \ref{tab6} and, apart from a slightly too low oscillation
amplitude for R Aqr, they fit well in the group of curves listed in Table \ref{tab3}, suggesting that binarity does not have a major impact on
the arguments developed in the present article.

X Oph is a Mira variable of spectral type Mno3-9 that had not been retained in Paper I for having a light curve too different
from the other Mno stars studied in the article. Indeed, its curve parameters listed in Table \ref{tab6} display much too low an oscillation
amplitude in comparison with the curves listed in Table \ref{tab1}. A likely explanation is the presence of a bright K-type giant companion
\citep{Fernie1959}, which, being unresolved, would be the source of the light detected in the minima of the light curve.

\subsection{Other stars}
In the present subsection we briefly comment about the light curves of other stars, which we came across on the occasion of the
present study and which may deserve some mention.

- R Car \citep{RosalesGuzman2023, RosalesGuzman2024} was recently the target of high resolution observations. The parameters
describing its light curves are listed in Table \ref{tab6}, compatible with both the Table \ref{tab1} and Table \ref{tab3} samples. Both in terms of luminosity,
temperature and size \citep{Takeuti2013} and in terms of light curve (Table \ref{tab6}), R Car is very similar to $o$ Cet suggesting that it has a
modest initial mass and is marginally dredging-up material from the helium burning shell.

- U Per is a Mira variable that was used by \citet{Cunha2020}, together with L$_2$ Pup, to illustrate general considerations that
they make on the stability of the pulsation regime of variable stars. The light curve of U Per (Table \ref{tab6}) belongs clearly to the Table 4
sample, close to the domain covered by carbon-rich stars.

- V Hya is a carbon star thought to be close to transiting to post-AGB \citep[][and references therein]{Planquart2024b}, with a high
mass-loss rate of 10$^{-5}$ M$_\odot$yr$^{-1}$ \citep{Knapp1999} and a mean spectral broadening of 13.5 \kms with a periodic variation
of 9 \kms\ suggesting rapid star rotation \citep{Barnbaum1995}. Numerous observations suggest the presence of an unseen companion
at $\sim$11 au distance causing periodic dimming ($\sim$17 years) of the light curve \citep[][and references therein]{Planquart2024a}. \citet{Sahai2022}
have used ALMA to detect the emission of CO lines with a good angular resolution and \citet{ZhaoGeisler2012} have used VLTI/MIDI
to observe the dust and molecular shells surrounding the star and compare their properties with those of four AGB stars
studied in the present article: R Aql, R Aqr, R Hya and W Hya. Both the periodic dimming and an insufficient coverage offered by
existing observations prevent a reliable evaluation of the parameters of the light curve, which is anyhow far from a regime of stable
pulsation. Yet, we selected cycles allowing for a crude estimate of their values, which are listed in Table \ref{tab6}. They fit well in the large
$W$ region of the Table \ref{tab5} sample.

- VX Sgr is a very massive and luminous oxygen-rich AGB star which has been the target of a multitude of observations
\citep[][and references therein]{Chiavassa2009, Tabernero2021}, which are still challenging interpretation. Its light curve is
outside the domain of stable pulsation and its complex nature prevents a meaningful evaluation of the curve parameters defined in the
present article. Yet, it does experience episodes of relatively stable pulsation, a particularly long one occurring between 1968 and
1997. During this episode, it is possible to estimate approximate values of the curve parameters, which are listed in Table \ref{tab6}. If it were
not for the period, which exceeds 2 years, larger than any of our sample, the other parameters would clearly assign the light curve to
Table \ref{tab4}, near the border with the Table \ref{tab5} curves of carbon-rich stars and significantly shifted in their direction.

\begin{deluxetable*}{lcl ccc ccc c}%{rrrrrrrrrrr}%
\tablenum{6}
\tablecaption{Parameters of the curves discussed in Section 4. In the case of the HBB and HBB-like curves, we apply different rules: we
mention $W_{\rm{1/5}}$ and $W_{\rm{R}}$ separately but refrain from quoting the value of their geometric mean; we do not quote values of $\varphi_{\rm{min}}$, which
would require to better specify its definition; we only quote values of $\Delta$ for R Cen, R Nor and AC Her, the irregularity of the other
curves preventing a meaningful definition; the parameters of R Cen are evaluated over the pre-1967 epoch; the parameters of SV Cas,
RU Cyg and DH Cyg are evaluated over episodes of relatively stable pulsation. \label{tab6}}
%\tablewidth{0pt}
\tablehead{
  \colhead{Name}&\colhead{$P$ (days)}&\colhead{Spctr}&\colhead{Tc}&\colhead{$^{12/13}$C}&\colhead{$C$ (mag)}&\colhead{$A$ (mag)}&\colhead{$W$}&\colhead{$\Delta$ (mag)}&\colhead{$\varphi_{\rm{min}}$}}
\startdata
\hline
\multicolumn{10}{c}{HBB and HBB-like}	\\
\hline
R Cen	&531	&M4-6	&no	&$-$	&2.2	&4.2	&0.34/0.63	&0.23	&$-$	\\
R Nor	&498	&M3-7	&no	&$-$	&2.3	&4.8	&0.28/0.53	&0.35	&$-$	\\
SV Cas	&450	&M5	&no	&$-$	&2.7	&2.2	&0.29/0.42	&$-$	&$-$	\\
RU Cyg	&468	&M9	&no	&32	&2.4	&1.7	&0.43/0.78	&$-$	&$-$	\\
DH Cyg	&531	&M6	&$-$	&$-$	&2.7	&2.8	&0.36/0.72	&$-$	&$-$	\\
AC Her	&75.4	&F4	&$-$	&$-$	&7	&1.1	&0.55/0.87	&0.08	&$-$	\\
\hline
\multicolumn{10}{c}{Mno outliers}\\
\hline
W Dra	&279	&M3-4	&no	&$-$	&2.7	&4.8	&0.35	&0.5	&0.59	\\
W Hya	&391	&M7.5-9	&no	&16	&0.4	&3.2	&0.37	&0.34	&0.48	\\
T Ari	&320	&M6-8	&no	&15	&1.6	&2.1	&0.5	&0.22	&0.55	\\
U CMi	&409	&M4	&no	&$-$	&2	&4	&0.47	&0.34	&0.39	\\
T UMi	&317	&M6	&no	&$-$	&1.4	&4.9	&0.21	&0.31	&0.53	\\
\hline
\multicolumn{10}{c}{Binaries}\\
\hline
R Aqr	&386	&M6.5-8.5	&yes	&16	&1.6	&4.2	&0.23	&0.6	&0.57	\\
$o$ Cet	&333	&M5-9	&yes	&10	&1.5	&5.6	&0.24	&0.62	&0.62	\\
X Oph	&334	&M3-9	&no	&12	&2	&1.7	&0.35	&0.22	&0.49	\\
\hline
\multicolumn{10}{c}{Other cases} \\
\hline
RU Tau	&589	&M3.5	&$-$	&$-$	&1.9	&3.7	&0.41	&$-$	&$-$	\\
U Ori	&372	&M6-9.5	&poss	&25	&2.6	&5.5	&0.24	&0.55	&0.61	\\
R Car	&308	&M5-8	&$-$	&$-$	&1.4	&5	&0.25	&0.47	&0.53	\\
U Per	&321	&M5.5-7	&prob	&$-$	&1.6	&3.2	&0.47	&0.22	&0.46	\\
V Hya	&530	&C6-7-N	&$-$	&$-$	&2.5	&2	&0.62	&0.25	&0.5	\\
VX Sgr	&759	&M8	&$-$	&$-$	&2	&4	&0.46	&0.56	&0.51	\\
\enddata
\end{deluxetable*}

\section{Discussion and conclusion}
\begin{figure*}
  \centering
  \includegraphics[width=7cm]{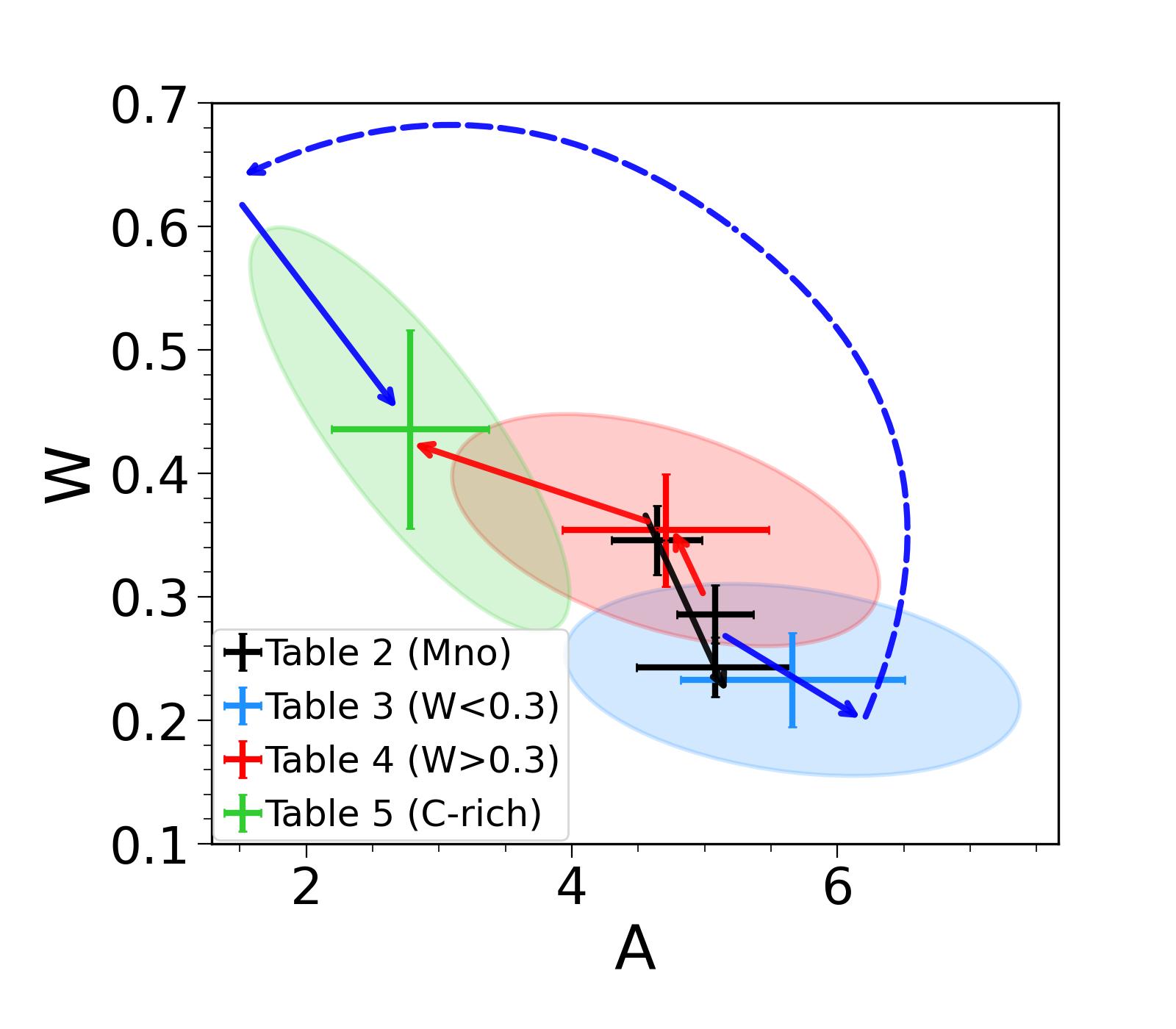}
  \includegraphics[width=7cm]{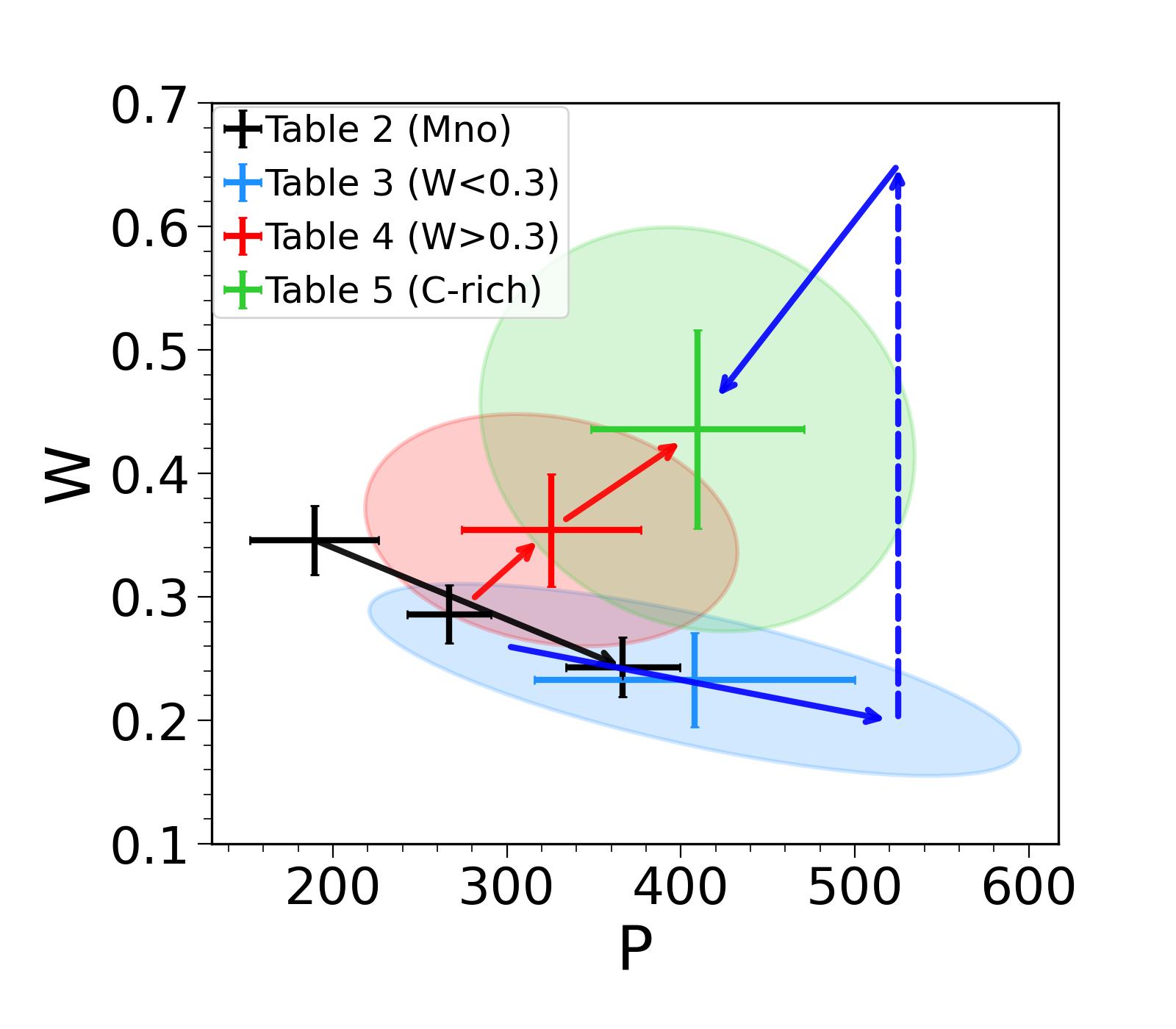}
  \caption{Schematic evolution of the light curves in parameter space ($W$ vs $A$ projection in the left panel and $W$ vs $P$ projection in the
right panel). Crosses show mean and rms values of the samples listed in the inserts. Coloured ellipses approximate the regions covered
by these samples. Arrows show the suggested evolution as described in the text: green for carbon stars, black for Mno stars of the
Table \ref{tab1} sample that will never experience a strong TDU event, blue and red for stars experiencing strong TDU events; the former are
expected to have lower initial masses than the latter and to possibly transit from oxygen-rich to carbon-rich via a chaotic pulsation
regime (dotted arrows).}
 \label{fig10}
\end{figure*}

\begin{figure*}
  \centering
  \includegraphics[width=4.4cm,trim={0cm 0 1.5cm 1cm},clip]{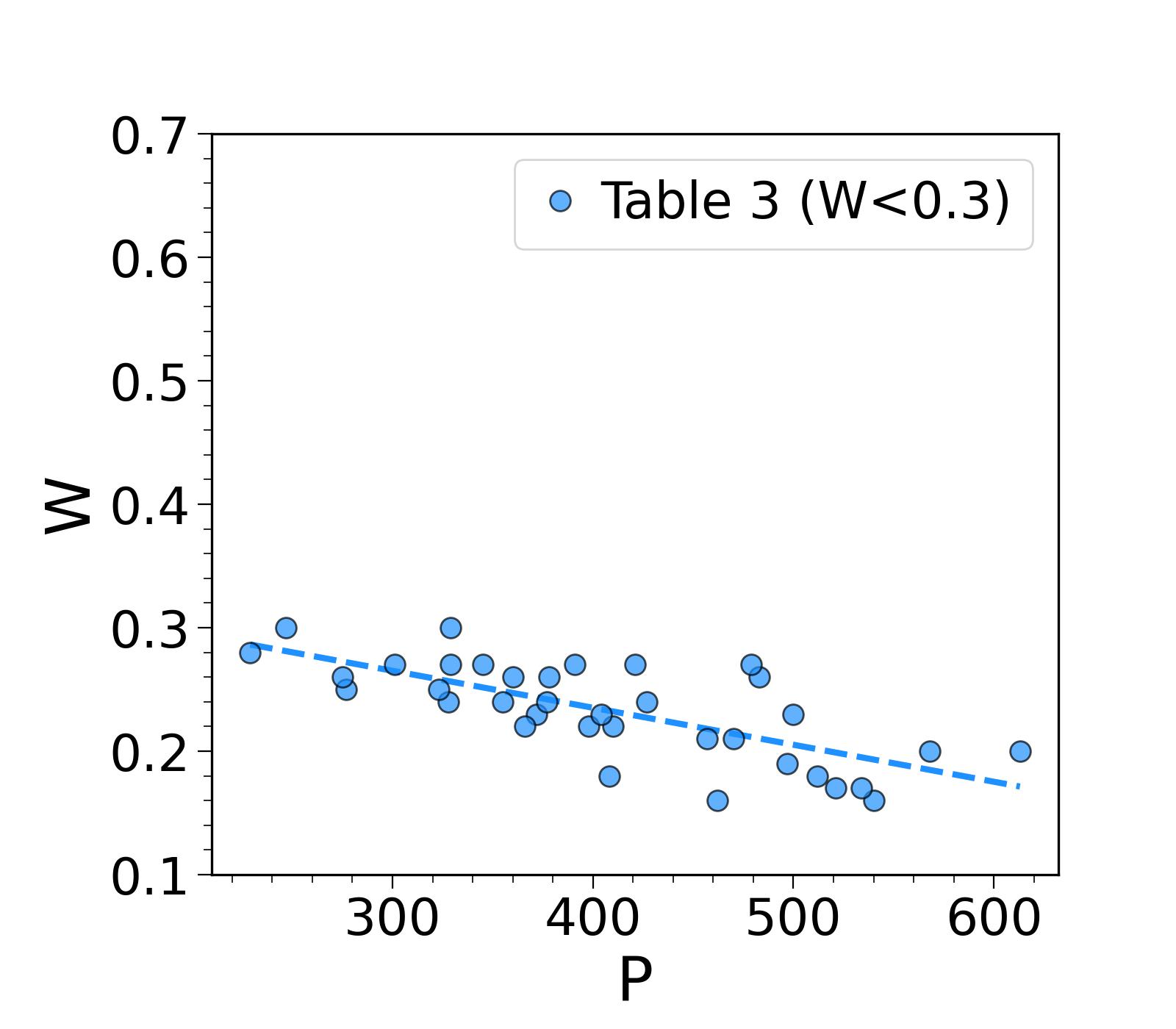}
  \includegraphics[width=4.4cm,trim={0cm 0 1.5cm 1cm},clip]{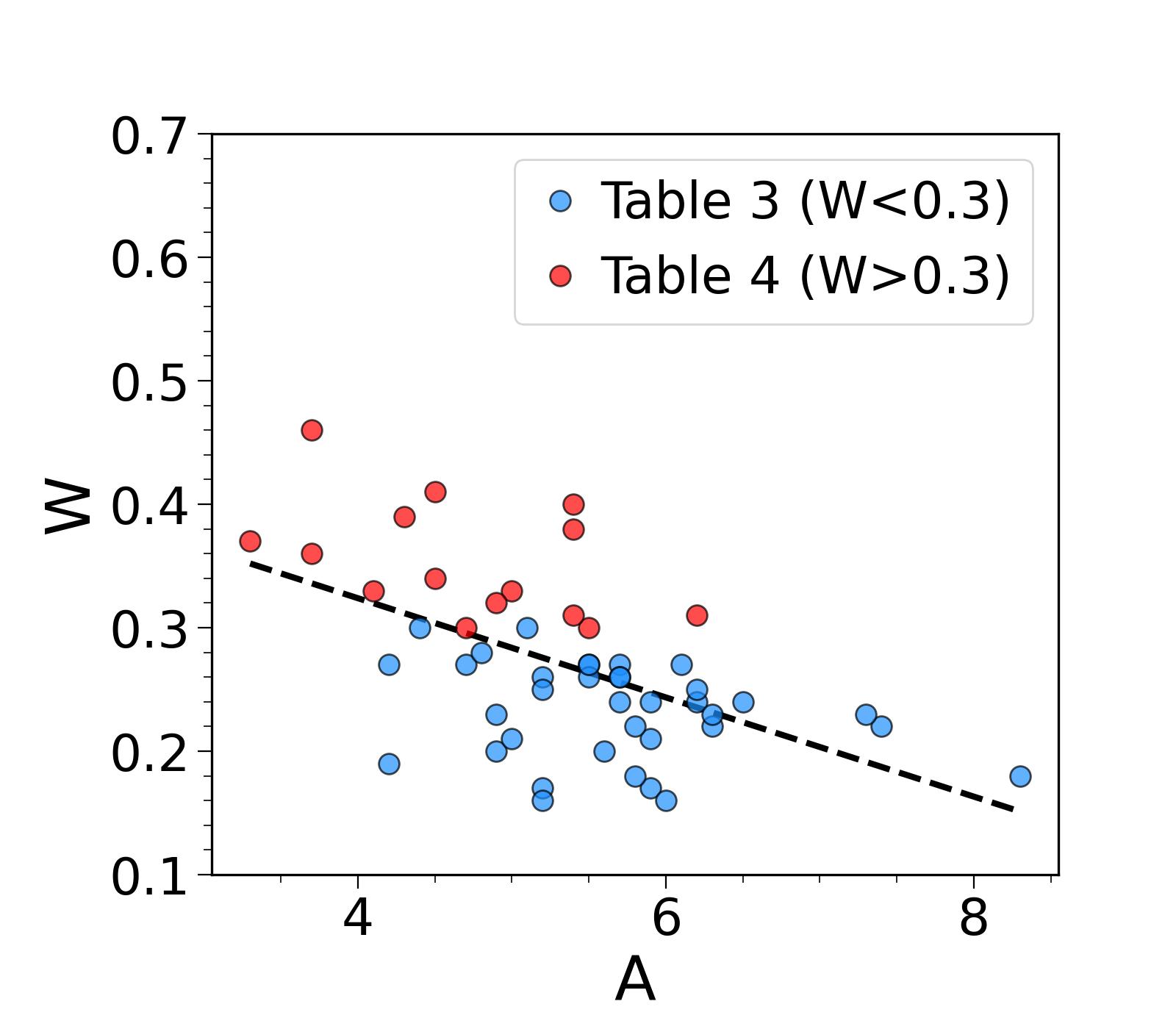}
  \includegraphics[width=4.4cm,trim={0cm 0 1.5cm 1cm},clip]{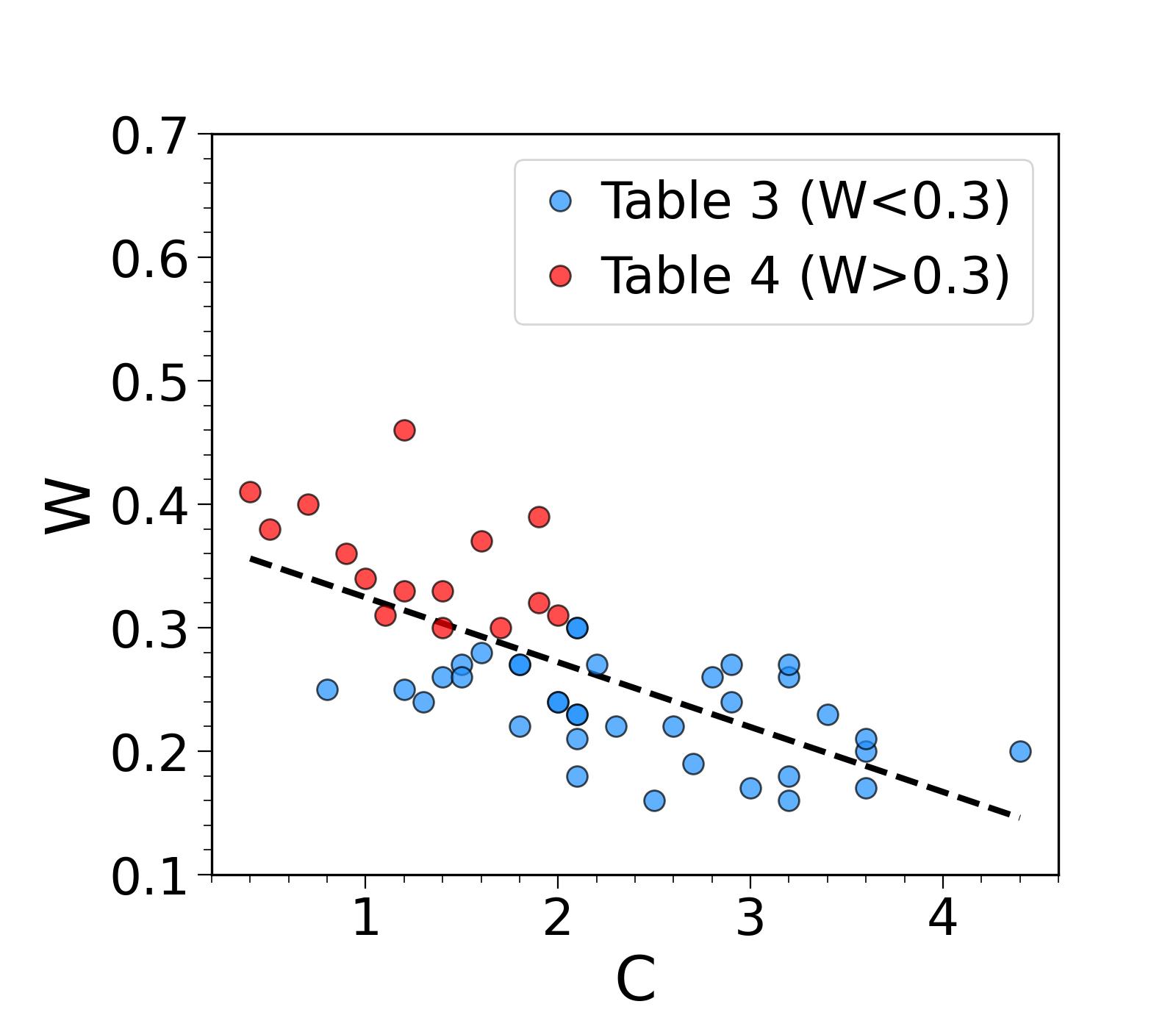}
  \includegraphics[width=4.4cm,trim={0cm 0 1.5cm 1cm},clip]{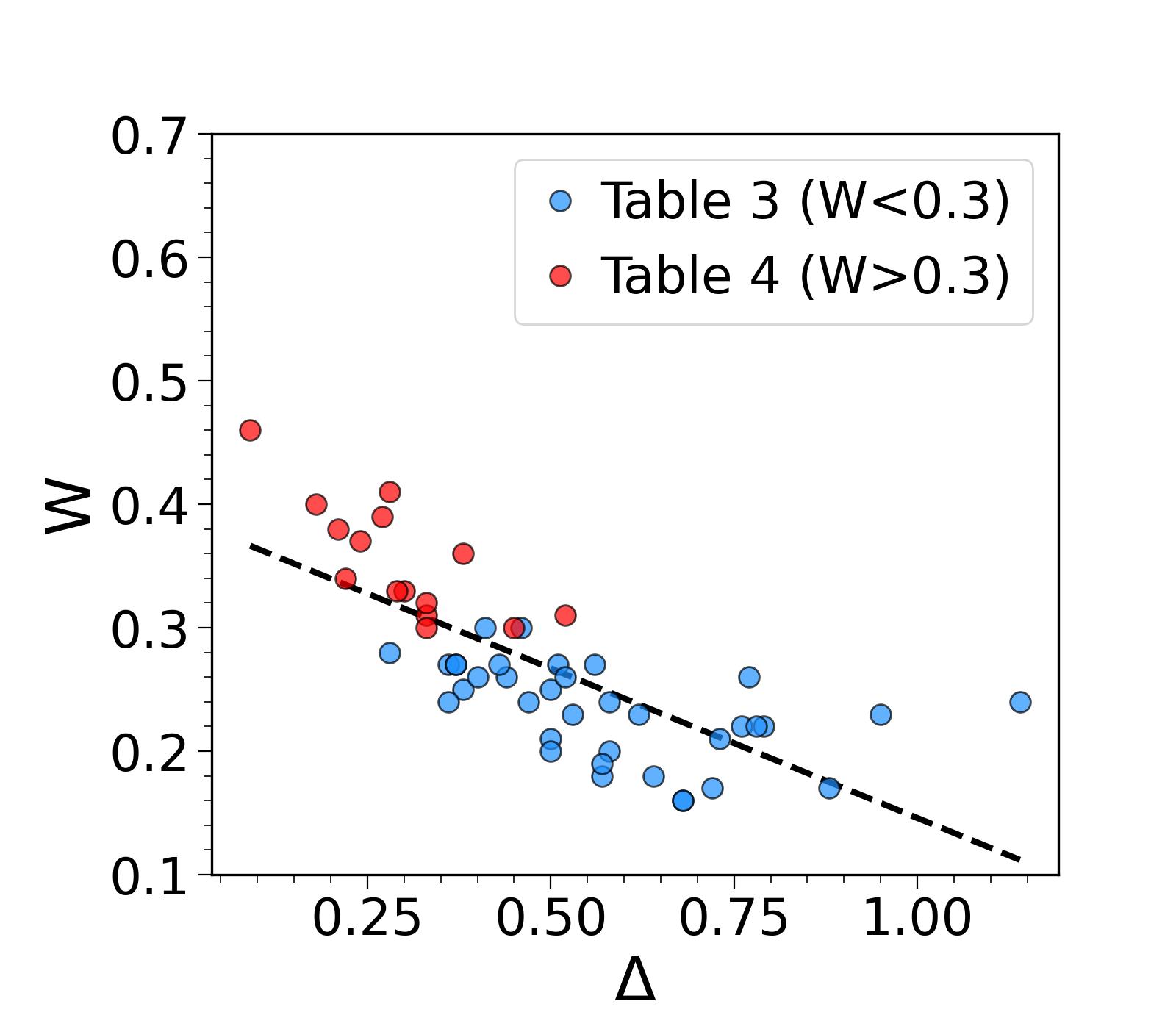}\\
  \includegraphics[width=4.4cm,trim={0cm 0 1.5cm 1cm},clip]{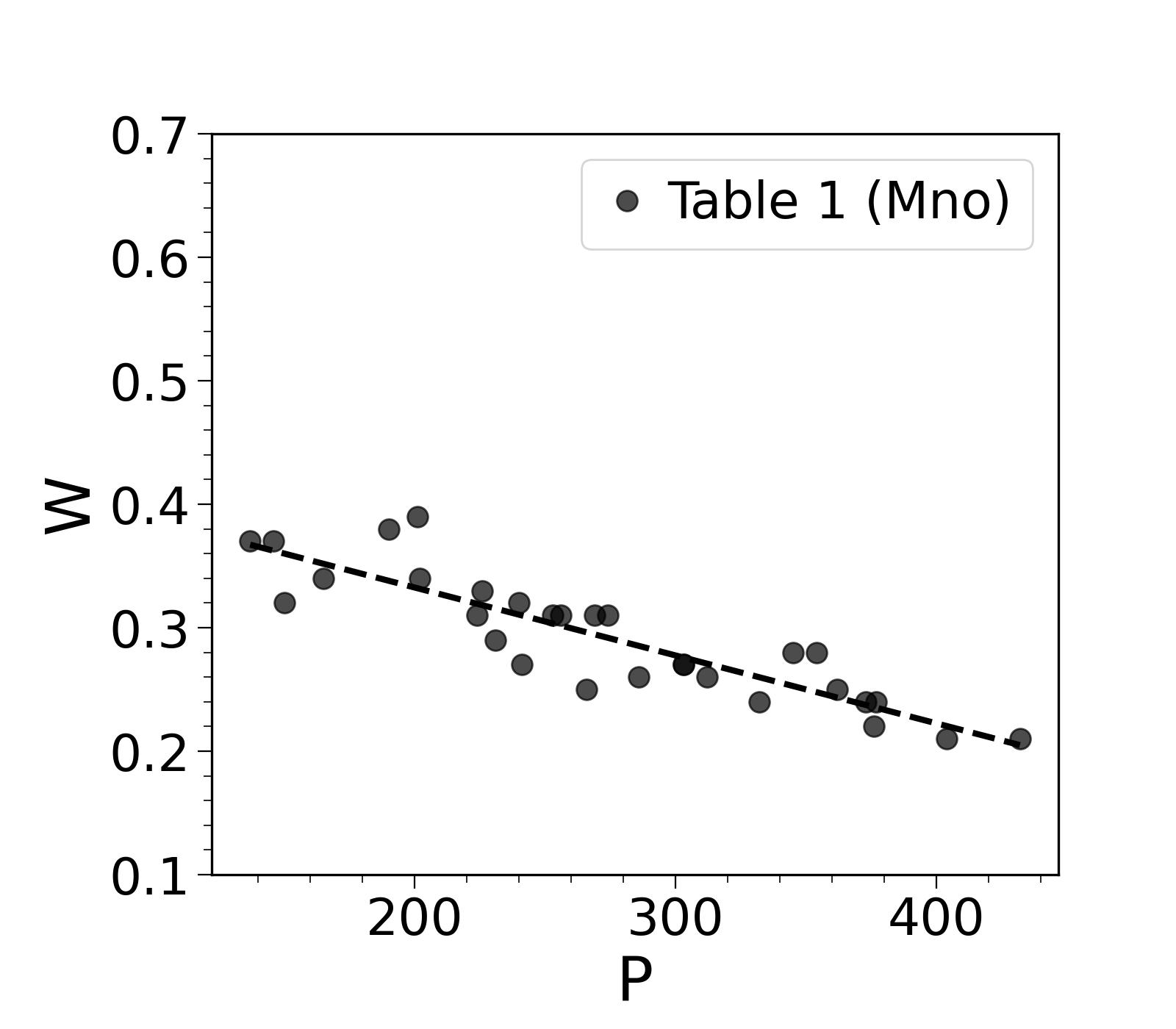}
  \includegraphics[width=4.4cm,trim={0cm 0 1.5cm 1cm},clip]{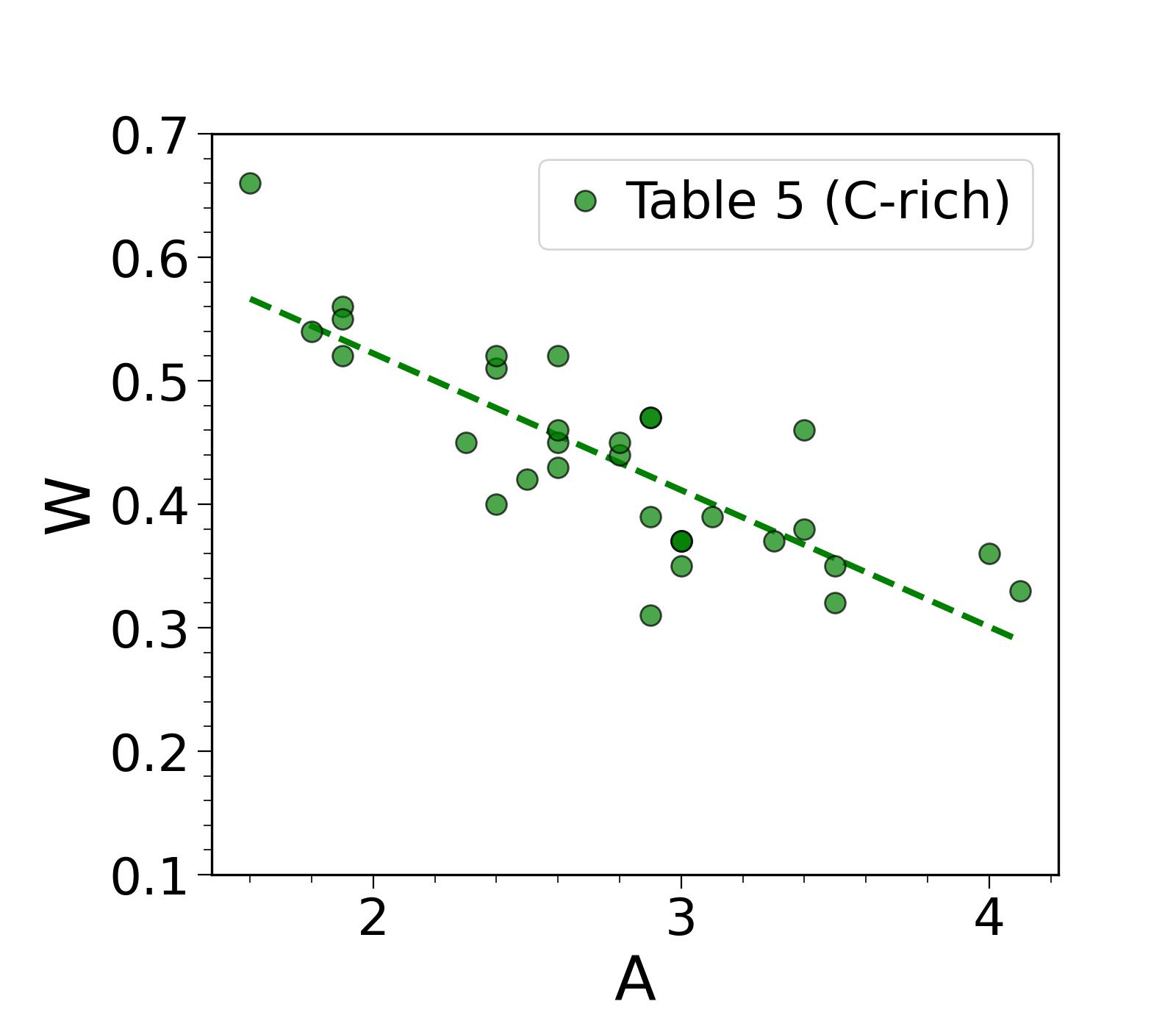}
  \includegraphics[width=4.4cm,trim={0cm 0 1.5cm 1cm},clip]{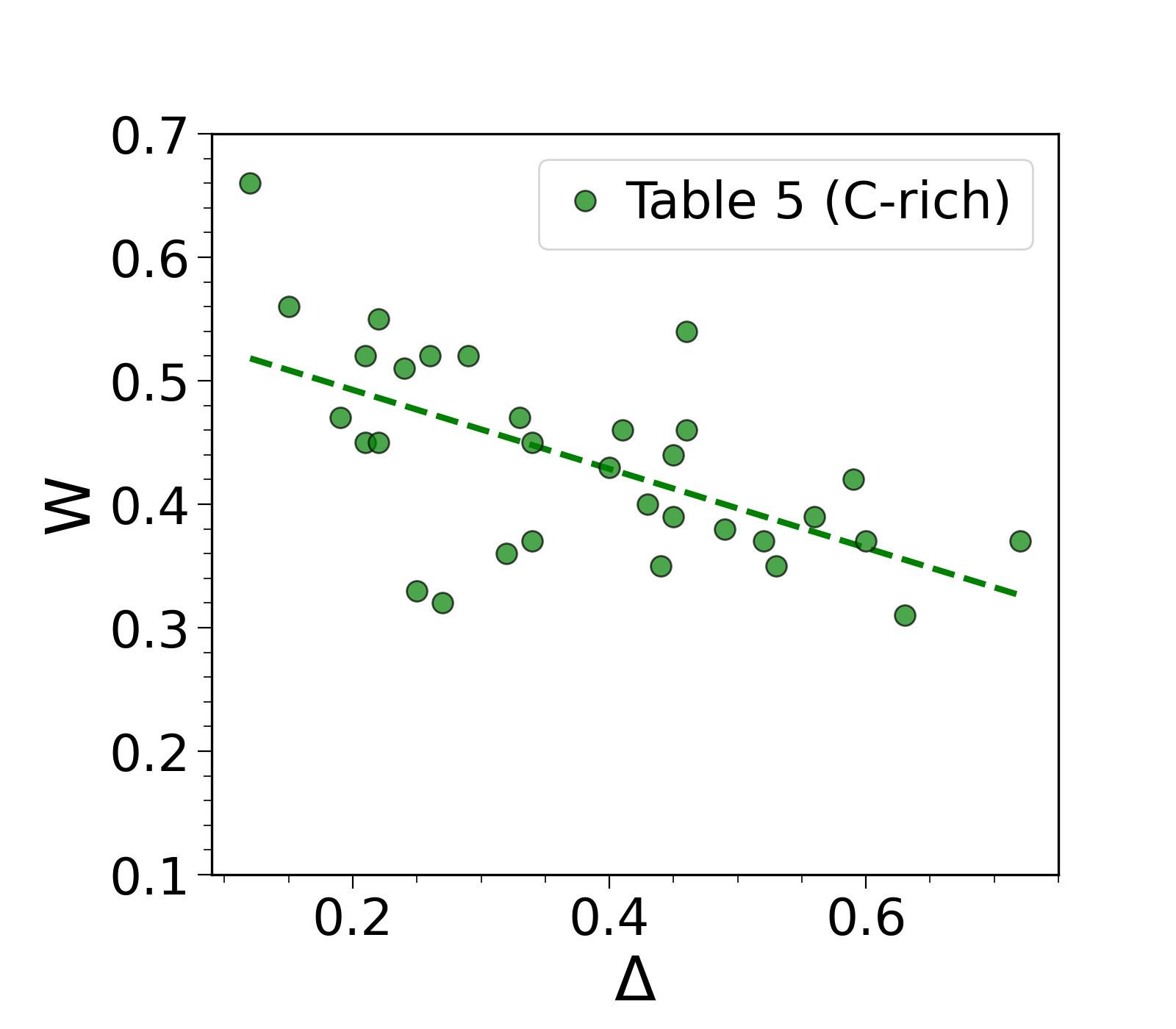}
  \includegraphics[width=4.4cm,trim={0cm 0 1.5cm 1cm},clip]{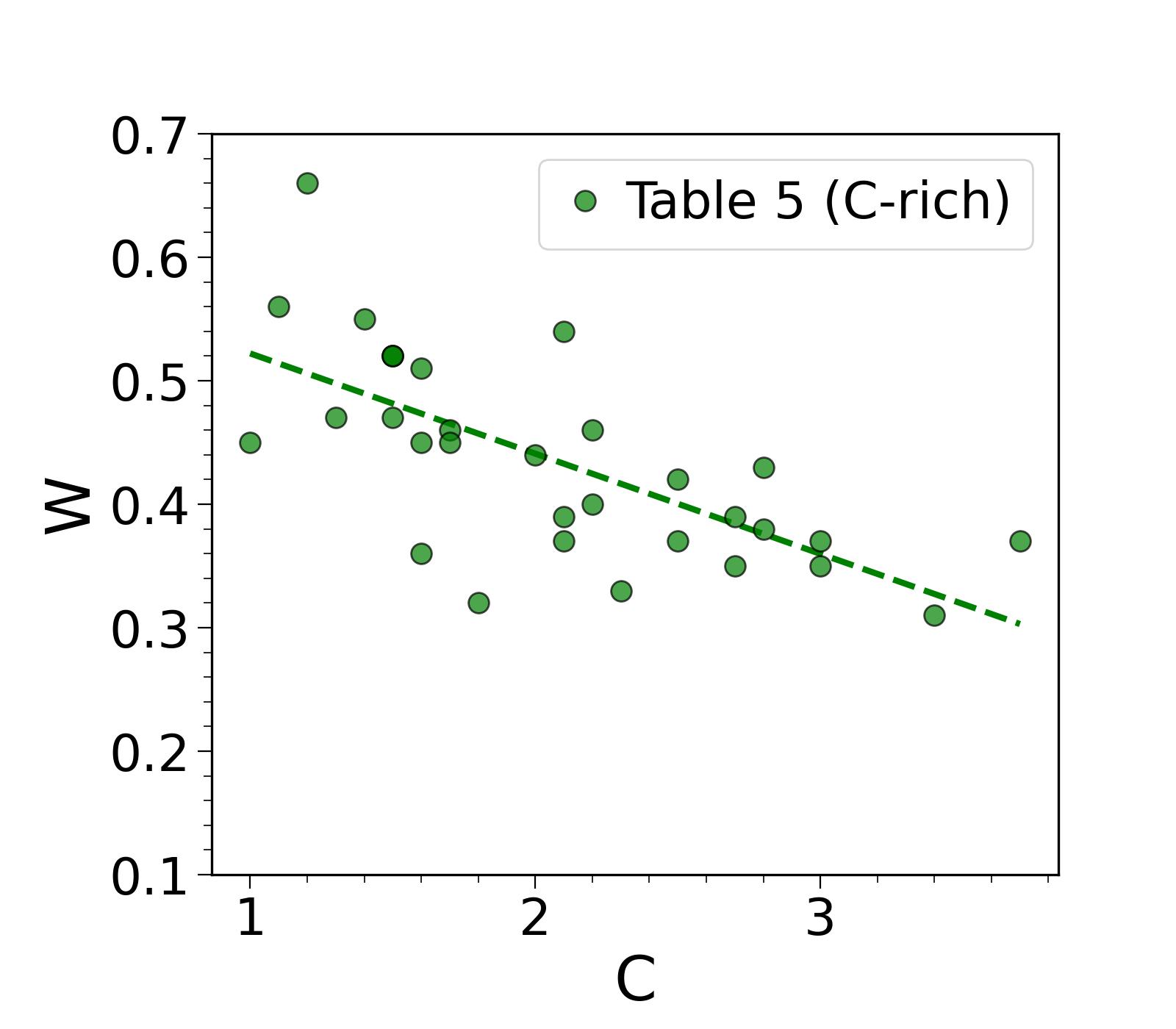}        
  \caption{Distributions of the curve parameters in projections on planes for which a significant correlation is observed. The lines are the results of the linear fits listed in Table \ref{tab8}.}
 \label{fig11}
\end{figure*}

The emphasis in the present article has been to try to reveal general relations between the evolution of stars along the AGB and the
parameters of their light curves, at the price of accepting some loss of rigour. For example, by using mean parameters $C$, $W$ and $\Delta$, we
lost the information contained in the differences between the members of the relevant pairs. Also, we accepted to ignore the
differences, often significant, between the uncertainties attached to the parameters that have been used. This must be kept in mind
when considering the results obtained and possibly extending the present work to future observations or analyses.
By using five parameters to characterize the main features of a light curve, we have been able to clarify the picture that had
been sketched in Paper I. Figure \ref{fig10} illustrates the main results, which we summarize below. Table \ref{tab7} lists, for each of the Table \ref{tab1}, \ref{tab3}, \ref{tab4} and \ref{tab5} samples,  the mean values and rms deviations with respect to the mean of the curve parameters. Table \ref{tab8} lists, for cases where a significant correlation has been observed between $W$ and another parameter, the result of a linear fit and the rms deviation $\Delta{W}$ of $W$ with respect to it. In three cases ($W$ vs $A$, $C$ and $\Delta$) the linear fit is made jointly to Table \ref{tab3} and Table \ref{tab4} curves. The linear fits are illustrated in Figure \ref{fig11}.

When on the early part of the AGB, stars expand, cool down and tend to pulsate with increasing periods; when their period
exceeds 100 days, they are classified as Mira variables and enter the sample considered in the present work. At that time, they are
oxygen-rich with an M spectral type and have not yet experienced sufficient TDU events for their impact to be detectable. The sample
of Mno curves listed in Table \ref{tab1} corresponds to such stars. As the temperature of the star decreases and as its size increases, the light
curve evolves from what was defined as first Mno family to what was defined as second Mno family in Paper I. Namely (black arrow
in Figure \ref{fig10}) the period increases, the width parameter $W$ decreases, the phase at minimal light increases, weak humps may appear on
the ascending branch close to minimal light, the pulsations become less regular, the colour indices increase and the ratio between the
oscillation amplitudes in the infrared and in the visible increases.

\begin{deluxetable*}{lcc ccc c}%{rrrrrrrrrrr}%
\tablenum{7}
\tablecaption{Mean values$\pm$rms deviations with respect to the mean of the curve parameters of the Table \ref{tab1}, \ref{tab3}, \ref{tab4} and \ref{tab5} samples. \label{tab7}}
%\tablewidth{0pt}
\tablehead{
  \colhead{}&\colhead{$P$ (days)}&\colhead{$W$}&\colhead{$A$ (mag)}&\colhead{$C$ (mag)}&\colhead{$\Delta$ (mag)}&\colhead{$\varphi_{\rm{min}}$}}
\startdata
Table 1	&274$\pm$79	&0.29$\pm$0.05	&4.9$\pm$0.5	&2.0$\pm$0.5	&0.44$\pm$0.11	&0.54$\pm$0.04	\\
Table 3	&410$\pm$92	&0.23$\pm$0.04	&5.8$\pm$0.8	&2.4$\pm$0.8	&0.56$\pm$0.20	&0.57$\pm$0.04	\\
Table 4	&326$\pm$52	&0.35$\pm$0.05	&4.7$\pm$0.8	&1.3$\pm$0.5	&0.29$\pm$0.10	&0.51$\pm$0.04	\\
Table 5	&410$\pm$61	&0.44$\pm$0.08	&2.8$\pm$0.6	&2.1$\pm$0.7	&0.39$\pm$0.15	&0.53$\pm$0.04	\\
\enddata
\end{deluxetable*}

\begin{deluxetable*}{lcc cc}%{rrrrrrrrrrr}%
\tablenum{8}
\tablecaption{Cases where a significant correlation has been observed between $W$ and another parameter, $\zeta$. The value of $\Delta{W}$,
  rms deviation of $W$ with respect to a linear fit of the form $W$$=$$a\zeta+b$, is listed in the fifth column. \label{tab8}}
\tablehead{
  \colhead{Parameter}&\colhead{Sample}&\colhead{a}&\colhead{b}&\colhead{$\Delta{W}$}}
\startdata
$W$ vs 10$^{-3}\times P$& Table 1&$-$0.550&0.443&0.023\\
$W$ vs 10$^{-3}\times P$& Table 3&$-$0.300&0.355&0.027\\
$W$ vs $A$&Tables 3+4&$-$0.040&0.485&0.057\\
$W$ vs $C$&Tables 3+4&$-$0.053&0.377&0.050\\
$W$ vs $\Delta$&Tables 3+4&$-$0.242&0.388&0.046\\
$W$ vs $A$&Table 5&$-$0.111&0.743&0.047\\
$W$ vs $C$&Table 5&$-$0.081&0.603&0.059\\
$W$ vs $\Delta$&Table 5&$-$0.320&0.557&0.064\\
\enddata
\end{deluxetable*}

What the present work has made clear is that the transition from Mno to Myes, namely the occurrence of TDU events strong
enough to have a detectable impact, occurs at very different stages along the TP-AGB. Indeed it is expected to never occur for stars of
low initial masses \citep{Herwig2005}, and the curves of such stars stay in the Table \ref{tab1} sample for the whole AGB life of the star. The
presence in the Table \ref{tab3} ($W$$<$0.3) sample of stars of Myes, and even S spectral types covering a broad spectrum of $W$ values makes a reliable interpretation in terms
  of their evolution along the AGB difficult. It suggests that TDU events may occur at very different stages on the TP-AGB; if such is the case, it would
  probably mean that they occur the earlier the larger the initial mass of the star, as shown by model simulations \citep{Herwig2005}. The curves of the Table \ref{tab3}
sample (blue arrow in Figure \ref{fig10}) seem to keep evolving in the same direction after the transition to the Myes and S spectral types: the
periods and amplitudes of the oscillations keep increasing and they become less regular. This is consistent with the expectation that
thermal pulses modify only progressively the internal state of the star. The absence of carbon-rich curves in the parameter space of the
curves close to the more evolved curves of the Table \ref{tab3} sample suggests that some of these curves leave the domain of stable pulsation
and become chaotic before the transition from oxygen-rich to carbon-rich.

Another major result obtained by the present work is the existence of curves leaving the Table \ref{tab1} sample very early and
transiting to the Myes and S spectral types with the width parameter $W$ increasing rather than decreasing. These are the curves of the
Table \ref{tab4} ($W$$>$0.3) sample and their location in the parameter space suggests that they will evolve to the Table \ref{tab5} sample when the stars will
become carbon-rich (red arrows in Figure \ref{fig10}). Here again, such a suggestion, albeit sensible and plausible, is far from being proven.
The dispersion of the curve parameters about their mean is large for both the Table \ref{tab4} and Table \ref{tab5} samples, showing the complexity of
the situation. In particular, the presence in the Table \ref{tab5} sample of curves evolving with decreasing $W$ (possible successors of the Table
\ref{tab3} sample) and of curves evolving with increasing $W$ (possible successors of the Table \ref{tab4} sample) contribute to such complexity. It is
natural to speculate that the stars of the Table \ref{tab4} sample have larger initial masses, and probably are closer to exhausting their
convective layers than the stars of the Table \ref{tab3} sample: the behaviour of the curves of HBB stars discussed in Subsection 4.1 gives
support to such speculation as do the curves of stars that are known to be close to post-AGB, such as V Hya and VX Sgr.

A major difficulty that has been recurrently met during the study is the lack of a clear picture of the physics governing the
transition from a stable to an unstable regime of pulsation, namely of the nature of the stability strip in the Hertzsprung-Russell
diagram. The distinction between Mira, semi-regular and irregular variables, like the distinction between sudden, continuous and
meandering period changes, are tools that prove useful in organizing the discussion, but they fail to provide a clear picture of the
physics at stake. An abundant literature is dedicated to what causes changes of pulsation regime, most of the time blaming the
occurrence of a thermal pulse, but little is known about how such changes actually develop. They clearly have to be seen in the
context of non-linear dynamics, with light curves prone to display features such as humps or double-peaking and to become chaotic,
but we are unable to answer simple questions such as: will R Dor, a nearby AGB star that has been extensively observed and studied,
see its light curve evolve from its current irregular regime \citep{Bedding1998} to a more stable one? and if yes, when will that happen?

We spotted a few cases that raise questions: DH Cyg seems to be a good HBB candidate and its spectrum deserves being
inspected accordingly. The light curve of W Hya suggests that the star is Tc-rich rather than Tc-poor, possibly revealing that it already
experienced significant TDU events, however not strong enough to produce a detectable technetium signal. T Ari is known to have a
Tc-poor spectrum but its light curve displays the features usually displayed by the curves of carbon-rich stars; together with U CMi, its
light curve profile has a very large $W$ parameter, challenging interpretation. T UMi is another well-known puzzling case of a Tc-poor
star and the inspection of its light curve confirms its peculiarity. Finally, inspection of its light curve has also confirmed the well-
known oddity of the behaviour of VX Sgr.

In summary, the present study has significantly clarified the picture that had been sketched in Paper I. In particular it has
offered a detailed description of the relation between stellar parameters and the properties of the light curves of the Table \ref{tab1} (Mno)
sample and it has revealed a clear distinction between the curves of the Table \ref{tab3} and Table \ref{tab4} samples, which had been previously
unnoticed. It has suggested that their evolution follows two different paths in the parameter space, the former possibly implying a
transition out of the regime of stable pulsation. However, it has raised many questions which remain unanswered and the
interpretations that it has suggested are far from being ascertained. It should serve as a guideline for future observations and analyses,
which would hopefully help with progressing with the general understanding of the issues that have been addressed.
In particular, a detailed study of the shape of the ascending branches of the light curves would probably help with clarifying some of the issues that are still unclear and the consideration of light curves from different sources, such as OGLE and ASAS-SN, would be welcome.
It makes one feel that light curves still carry messages and information that we have not yet been able to decrypt.

\section*{acknowledgments}
  
To the extent that the analyses presented in the present article may have contributed some progress, the credit belongs to the
innumerable observers around the world and to the AAVSO, without whom none of the results could have been obtained. We are
accordingly deeply indebted to them: to the many observers for the high quality of their observations and to the AAVSO for the
outstanding handling and reduction of the data that makes their use particularly easy and efficient. We very much thank the anonymous referee for pertinent comments that helped with improving the presentation of the article.
We are deeply grateful to Pr Tim Bedding for his interest in our work and for useful comments. Financial support from the World
Laboratory, the Rencontres du Vietnam and the Vietnam National Space Center is gratefully acknowledged. This research is funded by the Vietnam National
Foundation for Science and Technology Development (NAFOSTED) under grant number 103.99-2024.36.

\bibliography{lightcurve2}{}
\bibliographystyle{aasjournalv7}



\end{document}